\documentclass[twocolumn,showpacs,preprintnumbers,superscriptaddress,amsmath,amssymb,nofootinbib, floatfix]{revtex4}
\usepackage{graphicx}
\usepackage[section]{placeins}
\newcommand{\cd}{\makebox[0.08cm]{$\cdot$}}
\newcommand{\bg}[1]{\mbox{\boldmath $#1$}}
\newcommand{\sla}{\not\!}
\begin{document}
\vspace{0.5cm}
\title{Nonperturbative calculation of the anomalous magnetic moment \\ in the
Yukawa model within truncated Fock space}
\author{V.~A.~Karmanov}
\affiliation {Lebedev Physical Institute, Leninsky Prospekt 53,
119991 Moscow, Russia}
\author{J.-F.~Mathiot}
\affiliation {Clermont Universit\'e, Laboratoire de Physique
Corpusculaire, \\ BP10448, F-63000 Clermont-Ferrand, France}
\author{A.~V.~Smirnov}
\affiliation {Lebedev Physical Institute, Leninsky Prospekt 53,
119991 Moscow, Russia}

\bibliographystyle{unsrt}

\begin{abstract}
Within the covariant formulation of light-front dynamics, we
calculate the state vector of a physical fermion in the Yukawa
model. The state vector is decomposed in Fock sectors and we
consider the first three ones: the single constituent fermion, the
constituent fermion coupled to one scalar boson, and the
constituent fermion coupled to two scalar bosons. This last
three-body sector generates nontrivial and nonperturbative
contributions to the state vector, which are calculated
numerically. Field-theoretical divergences are regularized using
Pauli-Villars fermion and boson fields. Physical observables can
be unambiguously deduced using a systematic renormalization scheme
we have developed previously. As a  first application, we consider
the anomalous magnetic moment of the physical fermion.

\end{abstract}
\maketitle

\section{Introduction}\label{intro}

The understanding of hadronic systems in terms of their elementary
degrees  of freedom have been, and still is,  one of the most
challenging problems in particle and nuclear physics over the last
ten years. The phenomenological properties of hadrons are now
rather well understood in terms of models, like the constituent
quark model or the bag model. The understanding of their
properties from the original Lagrangian of QCD is however still
under active debate.

In nuclear physics, the properties of nuclear structure in terms
of  the exchanges of  pions
are also well known. They are described by
using a phenomenological nucleon-nucleon potential expressed in
terms of the exchanges of one pion,
two correlated pions and so on. However,
their complete description from an effective chiral Lagrangian is
still missing.

A common difficulty in both domains is the description of
relativistic bound systems. This description should be
nonperturbative from the start in order to be able to find, for
instance, the physical mass of the bound state from the pole of
the scattering amplitude. The problem is especially acute when
the interaction coupling constant is large.

One of the most relevant approaches aimed at studying relativistic
systems of interacting particles is light-front dynamics (LFD)
proposed initially by Dirac~\cite{Dirac}. LFD is a form of
Hamiltonian dynamics which deals with the state vector defined not
at a fixed time moment, but on the light front plane $t+z=0$, in its
traditional form. The state vector is then usually decomposed in a
series of Fock sectors, each containing a fixed number of
particles.

The use of LFD to investigate relativistic  bound
states has been advocated for a long time. However, while the
dynamics of few-body systems, based on a phenomenologically
constructed interaction, has developed rapidly, application of LFD
to field theory beyond a perturbative framework is
not yet under complete theoretical control.
This is due to the fact that any practical calculation relies on
taking into account only a restricted number of Fock sectors in
the state vector decomposition or, in other words, on the Fock
space truncation. This approximation strongly complicates the
renormalization procedure, in contrast to that in standard
perturbation theory. Indeed,  the full cancellation of
field-theoretical divergences which appear in a given Fock sector
requires taking into account contributions from other sectors. If
even a part of the latter is beyond our approximation, some
divergences may leave uncancelled. Mathematically, it reflects
itself in possible dependence of approximately calculated
observables on the regularization parameters (e.~g., cutoffs). This
prevents to make any physical predictions if we cannot control the
renormalization procedure in one way or another.

In a previous study~\cite{kms_08} (see also references therein) we
have developed an appropriate renormalization procedure -- the
so-called Fock sector dependent renormalization (FSDR) scheme --
in order to keep the cancellation of field-theoretical divergences
under permanent control. Our approach is based on the covariant
formulation of LFD (CLFD), where the state
vector is defined on an arbitrary light-front plane characterized
by a light-like four-vector $\omega$~\cite{cdkm} and given by the
equation $\omega\cd x = 0$. The covariant formulation is necessary
in order to control any violation of rotational invariance,
including that which is caused by the Fock space truncation. In
particular, this is important in order to formulate, in an
unambiguous way, the renormalization conditions one should impose
on the bare coupling constant (BCC) to relate it to the physical
one.

In Ref.~\cite{kms_08} we calculated the fermion state vector and
the electromagnetic form factors within the Yukawa model and QED
in the lowest nontrivial approximation, when the state vector
includes  only two Fock sectors given by one constituent fermion
and one constituent fermion coupled to one boson. For this
two-body Fock space truncation, the
electromagnetic form factors are identical to
those obtained in the second order of perturbation theory, giving
rise to a Schwinger-type correction to the fermion magnetic
moment. Note that this result is not surprising, in spite of the
fact we have not done any expansion in powers of the coupling
constant, since no other contributions to the fermion
electromagnetic vertex, apart from the perturbative ones (resummed
to all orders) is generated in the two-body truncation.

We shall present in this work the calculation of the fermion
anomalous magnetic moment (AMM) within the same
Yukawa model, but for the three-body Fock space truncation, when
the state vector includes an additional Fock sector containing one
constituent fermion coupled to two scalar bosons. The presence of
three-body states gives rise to nontrivial nonperturbative
contributions to the
AMM, which can not be fully incorporated in
perturbation theory. Besides that, the Yukawa model is a quite
nontrivial one from the point of view of the renormalization
procedure, since it exhibits simultaneously mass, vertex, and wave
function renormalization. 

The plan of the article is the following. We recall in
Sec.~\ref{bound} the main features of our nonperturbative approach
and the renormalization procedure. The eigenvalue equations are
derived in the three-body truncation in Sec.~\ref{three-body}. We
calculate the fermion electromagnetic form
factors in Sec.~\ref{elm} and present our numerical results for
the
AMM in Sec.~\ref{num}. Our conclusions are
drawn in Sec.~\ref{conc}. The appendices collect all necessary
details to calculate the
AMM.

\section{Bound state systems in light-front dynamics} \label{bound}
\subsection{General framework}

The state vector $\phi(p)$ describing any relativistic system with
total four-momentum $p$ forms a representation of the
Poincar\'e group. The four-momentum operator squared $\hat P^2$ is
one of the Casimir operators of this group and the state vector
satisfies the equation
\begin{equation}\label{eigen}
\hat P^2 \phi(p) = M^2  \phi(p),
\end{equation}
where $M$ is the mass of the physical system under consideration
and $p^2=M^2$.

LFD serves as an effective and convenient tool
to solve this eigenvalue equation. Indeed, one of
the main advantages of LFD is that, due to kinematical
constraints, the vacuum state of a system of interacting particles
coincides with the free vacuum, and all intermediate states result
from fluctuations of the physical system. One can thus construct
the state vector in terms of combinations of free fields, i.e.
decompose it in a series of Fock sectors:
\begin{equation}\label{Fock} \phi(p)=
\sum_{n=1}^{ \infty} \int dD_n \phi_n(k_1,\ldots,k_n;p)\left\vert n
\right>,
\end{equation}
where $\left\vert n \right>$ is the state containing $n$ free
particles with the four-momenta $k_1,\ldots,k_n$ and $\phi_n$'s
are relativistic $n$-body wave functions, the so-called Fock
components. The phase space volume element is represented
schematically by $d D_n$. In the following, we shall restrict our
study to a physical system composed of one fermion and $n-1$
bosons. In that case
\begin{equation}
\label{freefield}
\left\vert n \right> \equiv a^\dagger (k_1) c^\dagger(k_2) \ldots
c^\dagger(k_{n}) \left\vert 0 \right>,
\end{equation}
where $a^\dagger$ and $c^\dagger$ are fermion and boson creation
operators, respectively, and
\begin{equation}
\label{phipsi}
\phi_n(k_1,\ldots k_n;p)=\bar{u}(k_1)\psi_n(k_1,\ldots k_n;p)u(p),
\end{equation}
where $u$'s are bispinors. To completely determine the state
vector, we normalize it according to
\begin{equation}
\label{normep} \phi(p') ^\dagger \phi(p) = 2 p_0 \delta^{(3)}({\bf
p'}-{\bf p}).
\end{equation}
With the decomposition~(\ref{Fock}), the normalization
condition~(\ref{normep}) writes
\begin{equation}
\label{Inor} \sum_{n=1}^\infty I_n=1,
\end{equation}
where $I_n$ is the contribution of the $n$-body Fock sector to the
full norm. Explicit formula for $I_n$ can be found in
Ref.~\cite{kms_08}.

In the following, we shall use CLFD~\cite{cdkm}
as a general framework. The covariance of our approach is due to
the invariance of the light front plane equation. This implies
that $\omega$ is not the same in any reference frame, but varies
according to Lorentz transformations, like the coordinate $x$. It
is not the case in the standard formulation of LFD where $\omega$
is fixed to $\omega=(1,0,0,-1)$ in any reference frame.

The light-front momentum operator $\hat P_\rho$ can be
constructed from the energy-momentum tensor. It is decomposed
according to
\begin{equation}
\label{pshr}
\hat{P}_{\rho}=\hat{P}^{(0)}_{\rho}+\hat{P}^{int}_{\rho},
\end{equation}
where the two terms on the r.-h.~s. are, respectively, the free
(i.~e. independent of the coupling constant and counterterms) and
interaction parts of the four-momentum operator. The operator
$\hat{P}^{int}_{\rho}$ is related to the interaction Hamiltonian
$H^{int}(x)$ on the light front by
\begin{equation}
\label{pintham}
\hat{P}^{int}_{\rho}=\omega_{\rho}\int
H^{int}(x)\,\delta(\omega\cd x)\,d^4x.
\end{equation}

From the general transformation properties of the light-front
plane $\omega\cd x=0$, one can derive the following conservation
law~\cite{cdkm} for each Fock component:
\begin{equation}
\label{k1n}
k_1+k_2+\cdots +k_{n}=p+\omega\tau_n,
\end{equation}
where the quantity $\tau_n$  is a measure of how far the $n$-body
system is off the energy shell\footnote{The term "off the energy
shell" is borrowed from the equal-time dynamics where the spatial
components of the four-momenta are always conserved, but the
energies of intermediate states are not equal to the incoming
energy.}. It is completely determined by the conservation
law~(\ref{k1n}) and the on-mass shell condition for each
individual particle momentum:
\begin{equation}
\label{tau}
2 \omega \cd p \ \tau_n=(s_n-M^2),
\end{equation}
where $s_n=(k_1+\ldots +k_n)^2$.

\begin{figure}[ht!]
\begin{center}
\includegraphics[width=6.5cm]{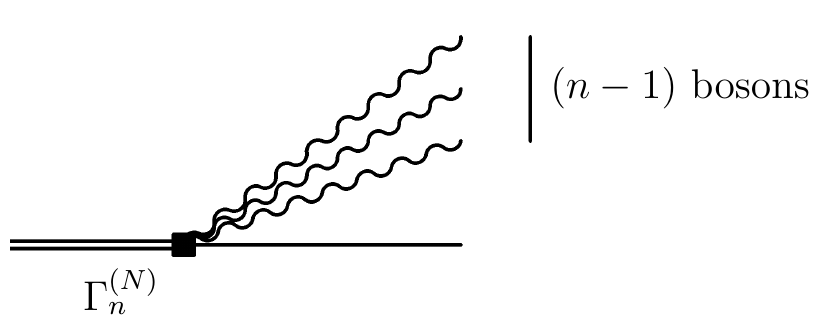}
\end{center}
 \caption{Vertex function of order $n$ 
 for the Fock space truncation of order $N$.}
 \label{gamman}
\end{figure}

It is convenient to introduce, instead of the wave functions
$\phi_n$, the vertex functions $\Gamma_n$ (which we will also
refer to as Fock components), defined by
\begin{equation}
\label{Gn}
\bar u(k_1) \Gamma_n u(p)=(s_n-M^2)\phi_n \equiv 2
\omega \cd p \ \tau_n \phi_n.
\end{equation}
The vertex function $\Gamma_n$ will be represented graphically by
the diagram of Fig.~\ref{gamman}. With the definition
\begin{equation}
\label{ga}
{\cal G}(p) = \sum_{n=1}^{ \infty} \int dD_n
\bar u(k_1) \Gamma_n(k_1,\ldots,k_n,p) u(p) \left\vert n \right>,
\end{equation}
the eigenvalue equation~(\ref{eigen}) writes~\cite{kms_08}
\begin{equation}
\label{gamma}
{\cal G}(p) = \frac{1}{2\pi}\int
\left[-\tilde{H}^{int}(\omega\tau)\right]\frac{d\tau}{\tau} {\cal
G}(p),
\end{equation}
where $\tilde H^{int}$ is the interaction Hamiltonian in momentum
space:
\begin{equation}
\label{eigenf}
\tilde{H}^{int}(p)=\int H^{int}(x)e^{-i p \cd x}d^4x.
\end{equation}

With the form~(\ref{gamma}), the eigenvalue equation can thus be
represented graphically, using the same rules as those derived in
Ref.~\cite{cdkm} for the calculation of matrix elements of the
$S$-matrix. This graph techniques was developed by
Kadyshevsky~\cite{kadysh} and transformed to the case of CLFD in
Ref.~\cite{karm76}.

The substitution of the decomposition~(\ref{ga}) into the
eigenvalue equation~(\ref{gamma}) results in an (infinite) system
of equations for the Fock components. In order to solve this
system in practice, we should make it finite, i.~e., truncate the
decomposition~(\ref{ga}), or equivalently~(\ref{Fock}),  by retaining only those Fock sectors
where the number of particles does not exceed some maximal value
$N$. The finite system can be solved numerically and
nonperturbatively, that is, for any value of the coupling
constant. This approach was developed in a series of
papers~\cite{bckm,kms2004,kms2007,kms_08}. In Ref.~\cite{kms_08}
it was applied to the case of the two-body truncation, i.~e. for
$N=2$.

\subsection{Renormalization conditions}
In order to be able to make definite predictions for physical
observables, one should also define a proper renormalization
scheme which allows to express, like in perturbation theory,
observables through the physical coupling constant  and masses and
exclude the bare ones. The basis of the state vector
decomposition, i.~e. the states $\left\vert n \right>$ in
Eq.~(\ref{freefield}) is constructed from free physical fermion
and boson fields, with their physical masses $m$ and $\mu$,
respectively. The interaction Hamiltonian contains the
corresponding mass counterterms (MC's) $\delta m$ and $\delta
\mu^2$ responsible for the fermion and boson mass renormalization.
Since we will not consider here the fluctuations of  the boson in
terms of fermion-antifermion pairs, we have to set $\delta \mu^2 =
0$. The MC $\delta m$ is determined from the eigenvalue
equation~(\ref{gamma}) by demanding that the bound state mass $M$
is equal, for the ground state, to the physical mass $m$ of the
constituent fermion. For this reason, we will distinguish $M$ and
$m$ only when it is necessary. Otherwise, we will set $M=m$.

Besides MC's, the interaction Hamiltonian includes also the BCC
$g_0$. The latter is determined, as in perturbation theory, by
relating the on-energy-shell two-body vertex function $\Gamma_2$
to the physical coupling constant $g$. As follows from
Eq.~(\ref{k1n}), taking $\Gamma_2$ on the energy shell is
equivalent to setting $\tau_2=0$. Once $M$ is identified with $m$,
the latter condition reduces to $s_2=(k_1+k_2)^2=m^2$ [see
Eq.~(\ref{tau})]. Below, for brevity, we will denote the
on-energy-shell two-body vertex function as $\Gamma_2(s_2=m^2)$ to
indicate that its arguments are connected by the corresponding
kinematical constraint.

Being a solution of the system of eigenvalue
equations~(\ref{gamma}), $\Gamma_2$ depends on the BCC $g_0$.
Hence, relating $\Gamma_2$ to $g$ is equivalent to relating $g_0$
to $g$, which just means
coupling constant renormalization.
This can be most easily done starting from the three point
function with all undressed on-mass-shell external lines, called
$\tilde \Gamma_2$. It is connected with the physical coupling
constant by the following standard relation (see e.g. Ref.~\cite{ps}):
\begin{equation}
\label{renor}
\sqrt{Z_f}  \ \tilde \Gamma_2(s_2=m^2) \sqrt{Z_f} \sqrt{Z_b}=g,
\end{equation}
where the $Z$-factors are the so-called field strength
renormalization  constants for the
fermion ($f$) (both in the initial and final state)
and boson ($b$) lines, respectively. This condition can also be
recovered by demanding that the residue of the fermion-boson
elastic scattering amplitude at $s_2=m^2$ equals $g^2$. One can
thus deduce the expression for $Z_f$ in terms of the full
fermion self-energy $\Sigma(\sla p)$:
\begin{equation}
\label{Zf} Z_f=\left[ 1- \left. \frac{\partial \Sigma({\sla
p})}{\partial  {\sla p}}\right\vert_{{p \!\!\!/ } = m}\right] ^{-1}
\end{equation}
and similarly for $Z_b$ as a function of the full boson
self-energy.

The two-body Fock component $\Gamma_2$ being a solution of the
eigenvalue equation~(\ref{gamma}) does not coincide with $\tilde
\Gamma_2$.  By definition, $\tilde \Gamma_2$ has no radiative
corrections to any of its three legs, while $\Gamma_2$, on the
contrary, includes such corrections. The relation between these
two vertex functions taken off the energy shell is rather
complicated in LFD. Fortunately, we need to know it on the energy
shell only, where it simplifies strongly, because the on-shell
radiative corrections mentioned above reduce to $c$-number
factors. Indeed, $\Gamma_2$ is a particular case of the general
vertex function shown in Fig.~\ref{gamman}, corresponding to
$n=2$, i.~e. two of its legs are represented by single external lines
(one for the constituent fermion and one for the constituent
boson), while the third leg, for the physical fermion, is shown by
a double line.  Radiative corrections to each of the two
external single lines are given by insertions of self-energy parts
with their subsequent summation. The latter, of course, can be
done directly within LFD by using the graph techniques rules, but
we will choose a simpler way.

Each on-energy-shell amplitude
calculated in LFD must coincide with that found in the standard
four-dimensional Feynman approach and taken on the mass shell.
Hence, $\Gamma_2(s_2=m^2)$ coincides with its on-mass-shell
Feynman counterpart \footnote{A three-leg vertex which enters, as
an internal sub-block, in physically observed amplitudes is always
off-shell. Taking it on shell, we imply its analytical
continuation into a nonphysical kinematical region}.

The summation
of radiative corrections to external lines in the Feynman approach
is technically easier than in LFD, since it can be done for each
of the two lines independently. We allow the external particle
momenta being off the mass shell (in order to avoid intermediate
singularities), then sum up the radiative corrections and finally
perform a limiting transition to the mass shell. Thus, summing a
chain series of self-energy blocks - together with 
the mass-counterterm insertion - on the constituent fermionic
line with the four-momentum $k_1$ brings the factor
\newpage
\begin{widetext}
\begin{eqnarray}
\lim_{{k \!\!\! /}_1\to m}\left[1+ \frac{\Sigma_r({\sla
k}_1)}{{\sla k}_1-m}+\frac{\Sigma_r({\sla k}_1)}
{{\sla k}_1-m} \times\frac{\Sigma_r({\sla
k}_1)}{{\sla k}_1-m}+\ldots\right] &=&\lim_{{k \!\!\! /}_1\to m}\left(\frac{{\sla k}_1-m}{{\sla
k}_1-m-\Sigma_r({\sla k}_1)}\right) \nonumber \\
&=& \left[ 1- \left. \frac{\partial \Sigma_r({\sla
k}_1)}{\partial  {\sla k}_1}\right\vert_{{k \!\!\! /}_1= m}\right]
^{-1},\label{calZ}
\end{eqnarray}
\end{widetext}
where $\Sigma_r({\sla k}_1)=\Sigma({\sla k}_1)-\Sigma(m)$.
This factor is nothing else
than $Z_f$ given by Eq.~(\ref{Zf}). 
Analogous procedure for the
boson line leads to the factor $Z_b$. The total factor which
appears due to the dressing of the two
constituent lines is therefore $Z_fZ_b$.

Concerning the double fermionic line in $\Gamma_2$, its renormalization
factor is related to the normalization condition
for the state vector. The eigenvalue equation~(\ref{gamma})
transforms into a homogeneous system of linear integral equations
for the vertex functions. Hence, the solution is determined up to
an arbitrary common factor. In practice, for solving this system
of equations, it is convenient to fix the (constant) one-body Fock
component $\phi_1$, so that the other components ($\Gamma_2$,
$\Gamma_3$, etc.) become proportional to it.  Then the double line
in $\Gamma_2$ brings the factor
$\phi_1$ which, in its turn, is
determined by the normalization condition~(\ref{normep}) for the
state vector.
Since $\phi_1^2$ is just the norm of the one-body
Fock sector, $\phi_1$ is equal to $\sqrt{I_1}$, where $I_1$ is the first term in
the sum~(\ref{Inor}).

The relation between $\Gamma_2$ and
$\tilde{\Gamma}_2$ becomes therefore on the energy shell
\begin{equation}
\label{G2on}
\Gamma_2(s_2=m^2) = \sqrt{I_1} \ \tilde \Gamma_2(s_2=m^2) \ Z_f
Z_b.
\end{equation}
Excluding $\tilde \Gamma_2(s_2=m^2)$ from Eqs.~(\ref{renor})
and~(\ref{G2on}), we find  that the renormalization condition
reads
\begin{equation}
\label{Gam3a}
\Gamma_2(s_2=m^2)=g \sqrt{I_1} \sqrt{Z_b},
\end{equation}
where  $\Gamma_2$ is expressed from the eigenvalue
equation~(\ref{gamma}), through the BCC $g_0$.
A similar discussion  of the renormalization condition in terms of the  one-body component $\phi_1$ was already done in Ref.~\cite{hb_99}.

When we neglect  the boson dressing by fermion-antifermion
fluctuations, as we do in this work, the condition~(\ref{Gam3a})
finally reduces to
\begin{equation}
\label{renorN1}
\Gamma_2 (s_2=m^2) =  g \sqrt{I_1}.
\end{equation}
Note that there is a simple relation
between the one-body normalization factor $I_1$ and the field
strength renormalization factor $Z_f$:
\begin{equation}
\label{IZ}
Z_f =I_1,
\end{equation}
as shown in Appendix~\ref{renorma}. 

\subsection{Renormalization scheme}
The above conditions imposed on the BCC and MC are necessary in
order to express physical observables, like the electromagnetic
form factors, through the measurable coupling constant and masses.
As a consequence, one should expect full cancellation of
divergences.

\begin{figure}[btph]
\begin{center}
\includegraphics[width=20pc]{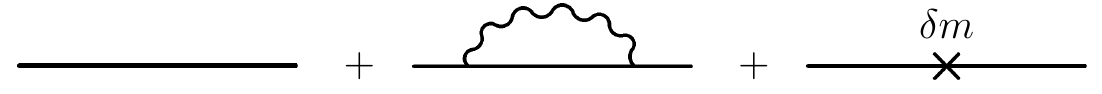}
\caption{Renormalization of the fermion propagator in
the second order of perturbation
theory.\label{self}}
\end{center}
\end{figure}

Such a program could be realized in perturbation theory or
nonperturbatively if the Fock space is not truncated. The latter case
is hardly achieved in practice. Usually, Fock space is truncated
to a finite order $N$ of admitted Fock sectors, and the
cancellation of divergences is not anymore guaranteed. For
instance, looking at Fig.~\ref{self} for the calculation of the
fermion propagator in the second order of perturbation theory, one
immediately realizes that the cancellation of divergences between
the self-energy contribution (of order two in the Fock
decomposition) and the fermion MC (of order one) involves two
different Fock sectors~\cite{kms_08}.
This means that, as a necessary condition for
the cancellation of
divergences, any MC and, more generally, any BCC should be
associated  with the number of particles present (or ``in
flight'') in a given Fock sector. In other words, all MC's and
BCC's must depend on the Fock sector under consideration. The
original MC, $\delta m$, and the fermion-boson BCC, $g_0$, should
thus be extended each to a whole series:
\begin{subequations}
\label{fsdr}
\begin{eqnarray}
\label{g0l} g_0 &\to&  g_{0l}\ ,\\
\label{dml} \delta m  &\to& \delta m_l,
\end{eqnarray}
\end{subequations}
with $l=1,2,\ldots N$. The quantities $g_{0l}$ and $\delta m_l$
are calculated by solving the systems of equations for the
vertex functions in the $N=1$, $N=2$, $N=3$, ... approximations
successively. This FSDR scheme has been  proposed initially in
Ref.~\cite{wp} and developed as a full renormalization scheme in
Ref.~\cite{kms_08}. An alternative approach, also
in the Pauli-Villars (PV) regulated Yukawa model with the two-boson
truncation, but with a sector-independent renormalization scheme, was
developed in Ref.~\cite{BHMc}.

Note that the series~(\ref{fsdr}) does not imply that we have an
infinite  number of counterterms or bare parameters. We still have
the original ones $g_0$ and $\delta m$ in the Hamiltonian we start
with, but they have different values according to the level of
approximation used in the calculation. In the limit of an infinite
$N$, and if the Fock sector expansion converges, $g_{0N}$ and
$\delta m_N$ should turn to the true BCC and the MC, respectively.
This is completely analogous to the case of perturbation theory
where, at each order $n$, one determines $g_0^{(n)}$ and $\delta
m^{(n)}$.

Apart from the mass and
vertex radiative corrections, the third type of divergences arises
from the field renormalization, i.~e. from the constants $Z_f$ and
$Z_b$.  The values of these constants should also depend on the
maximal number $N$ of particles kept in a given truncation.
Consider for instance the vertex function $\Gamma_2$ represented by
Fig.~\ref{gamman} for $n=2$. The dressing of the physical fermion
leg (the factor $\sqrt{I_1}$) should be calculated for the
truncation to the $N$-th order. The situation changes however for
the constituent (single) fermion line. The state in which the
constituent fermion is considered already contains one constituent
boson. Hence, even if the boson line is not dressed, the dressing
of the constituent fermion leg involves radiative corrections of
order $(N-1)$. In other words, the dressing factor $Z_f$ for the
constituent fermion leg must be calculated for the lower,
$(N-1)$-body truncation. Otherwise, we would go beyond our
approximation, since the effective number of particles in which
the physical fermion can fluctuate would exceed $N$.

Taking this into account, the relations~(\ref{renor})
and~(\ref{G2on}) for a finite order truncation $N$ (and in the
absence of boson dressing) obtain the following form:
\begin{subequations}
\begin{eqnarray}
\label{renort}
&&\sqrt{Z^{(N)}_f}   \tilde \Gamma_2(s_2=m^2) \sqrt{Z^{(N-1)}_f}
=g,\\
&&\Gamma_2(s_2=m^2) = \sqrt{I^{(N)}_1}  \tilde \Gamma_2(s_2=m^2)
Z_f^{(N-1)}. \nonumber \\
\end{eqnarray}
\end{subequations}
The superscripts $(N)$ and $(N-1)$ here and below just indicate
the order of the Fock space truncation in which the corresponding
quantities are calculated. 

It follows from Eqs.~(\ref{renort}) and~(\ref{IZ}) that the
renormalization condition~(\ref{renorN1}) simply writes
\begin{equation}
\label{Gam3c}
\Gamma_2(s_2=m^2)= g \sqrt{I_1^{(N-1)}},
\end{equation}
in the absence of boson dressing.

For the simplest case of the two-body truncation, $N=2$, one thus gets
\begin{equation}
\label{rencond2}
\Gamma_2 ^{(2)}(s_2=m^2) =  g,
\end{equation}
since $I_1^{(1)}=1$. We recover here the
condition given in Ref.~\cite{kms_08}. This condition is however
valid only for $N=2$.

\section{Yukawa model in three-body truncated Fock space}
\label{three-body}
\subsection{Eigenvalue equations} \label{Eigen}
We consider in this study the Yukawa model: a spin-1/2 fermion
interacting with massive spinless bosons. The regularization is
provided by the
PV method. In addition to physical particles, we introduce therefore one PV fermion and one PV boson
with (large) masses $m_1$ and $\mu_1$, respectively. This amounts
to extend the physical Fock space to embrace negatively normalized PV
particles~\cite{kms_08}. The interaction Hamiltonian in
Eq.~(\ref{gamma}) is given by
\begin{equation*}
{H}^{int}=-g_{0}\bar{\psi}'\psi' \varphi'-\delta m
\bar{\psi}'\psi',
\end{equation*}
with
\begin{equation}
\psi'=\psi+\psi_{PV},\quad \varphi'=\varphi+\varphi_{PV},
\end{equation}
where $\psi$ and $\varphi$ are the free physical fermion and boson
field operators, while $\psi_{PV}$ and $\varphi_{PV}$ are their PV
partners, with negative norm. The bosons are supposed to be
neutral.

\begin{figure}[ht!]
\begin{center}
\includegraphics[width=8.5cm]{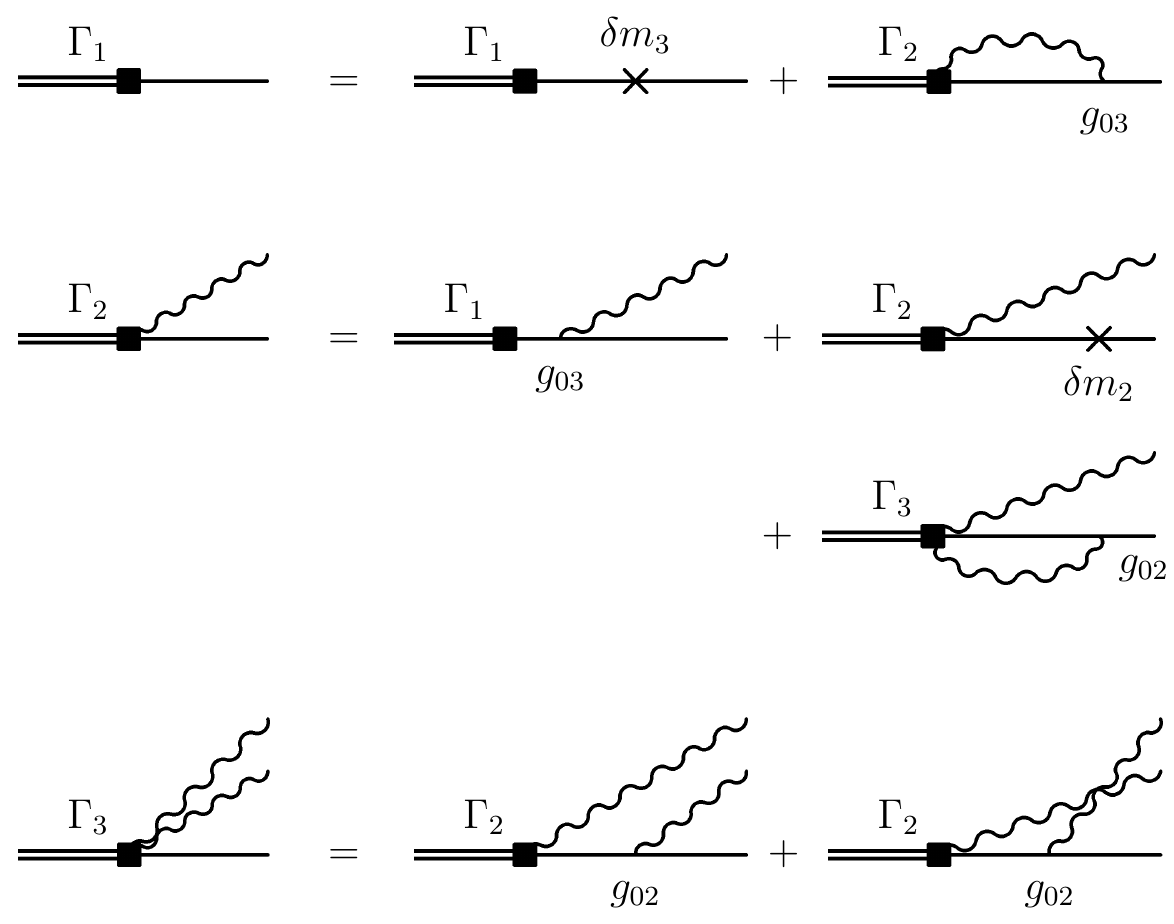}
\end{center}
 \caption{System of equations for the vertex functions in
the case of three-body Fock space
truncation.}\label{syst_eq}
\end{figure}

In the three-body truncation, the system of coupled equations for
the vertex functions, derived from the eigenvalue
equation~(\ref{gamma}), is shown graphically in
Fig.~\ref{syst_eq}.
On the r.-h.~s. of the last equation,
the sum of the diagrams with
permutated boson legs appears, reflecting the symmetrization of
the amplitude due to the identity of bosons. Expressing $\Gamma_3$ through $\Gamma_2$ by means
of this equation, and substituting
the result into the second equation, we can exclude the highest
order vertex function $\Gamma_3$ from the full system of
equations. We thus obtain a reduced equation for the two-body
vertex function.
\begin{figure}[ht!]
\begin{center}
\includegraphics[width=8.5cm]{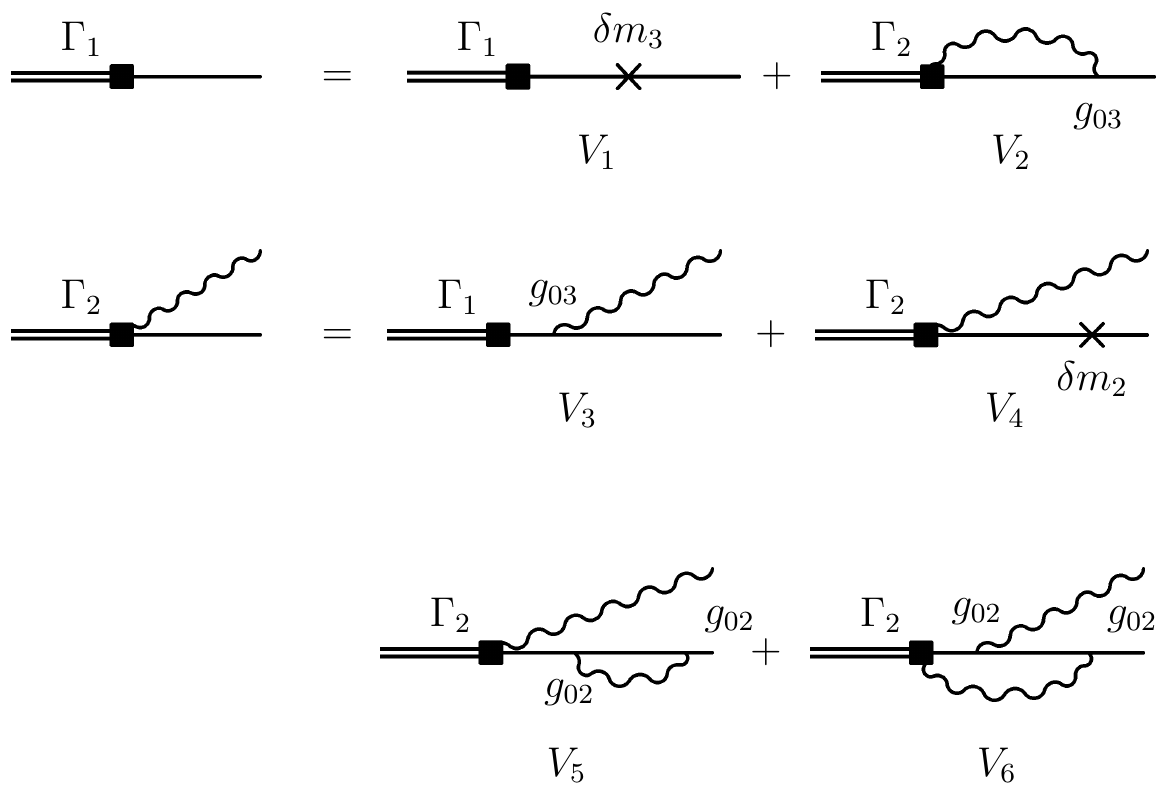}
\caption{Reduced equation for the two-body vertex function,
obtained from that shown in Fig.~\protect{\ref{syst_eq}} after the
exclusion of the three-body component.\label{syst_eq_r}}
\end{center}
\end{figure}
Together with the first equation in Fig.~\ref{syst_eq}, it forms a
system of equations involving the vertex functions $\Gamma_{1}$
and $\Gamma_{2}$ only, as shown in Fig.~\ref{syst_eq_r}. Analytically, these equations read
\begin{subequations}
\label{eq245}
\begin{eqnarray}
\label{eq24}
\bar{u}(p_{1i})\Gamma_{1}^{i}u(p) &=&
\bar{u}(p_{1i})\left(V_{1}+V_{2}\right)u(p),\\
\bar{u}(k_{1i})\Gamma_{2}^{ij}u(p) &=&
\bar{u}(k_{1i})\left(V_{3}+V_{45}+V_{6}\right)u(p), \nonumber \\
\label{eq25}
\end{eqnarray}
\end{subequations}
where the indices $i$ and $j$ refer to whether the line of a
constituent fermion ($i$) or a constituent boson ($j$) corresponds
to a physical particle ($i,j=0$) or to a PV one ($i,j=1$).
The term $V_{45}$ means the sum of $V_4$ and $V_5$. The
explicit expressions for $V_{1-6}$ are given in
Appendix~\ref{V16}. Note that the first equation in
Fig.~\ref{syst_eq} is just a constraint which determines $\delta
m_3$. The contribution $V_5$ on the r.-h.~s. of
Fig.~\ref{syst_eq_r} involves the two-body self-energy depending,
due to the departure off the energy shell, on a four-momentum $k$ with
$k^2\neq m^2$.This self-energy is decomposed, in CLFD, according
to~\cite{kms2007}:
\begin{equation}
\label{Sigk} \Sigma({\sla k}) = g_{02}^2 \left[ {\cal
A}(k^2)+{\cal B}(k^2) \frac{\sla k}{m}+{\cal C}(k^2)\frac{m {\sla
\omega}}{\omega \cd k} \right],
\end{equation}
where the factors $m$ are here for convenience only. The
coefficients ${\cal A}$, ${\cal B}$, and ${\cal C}$ are calculated
in Appendix~\ref{secoef}.

\begin{figure}[ht!]
\begin{center}
\includegraphics[width=6cm]{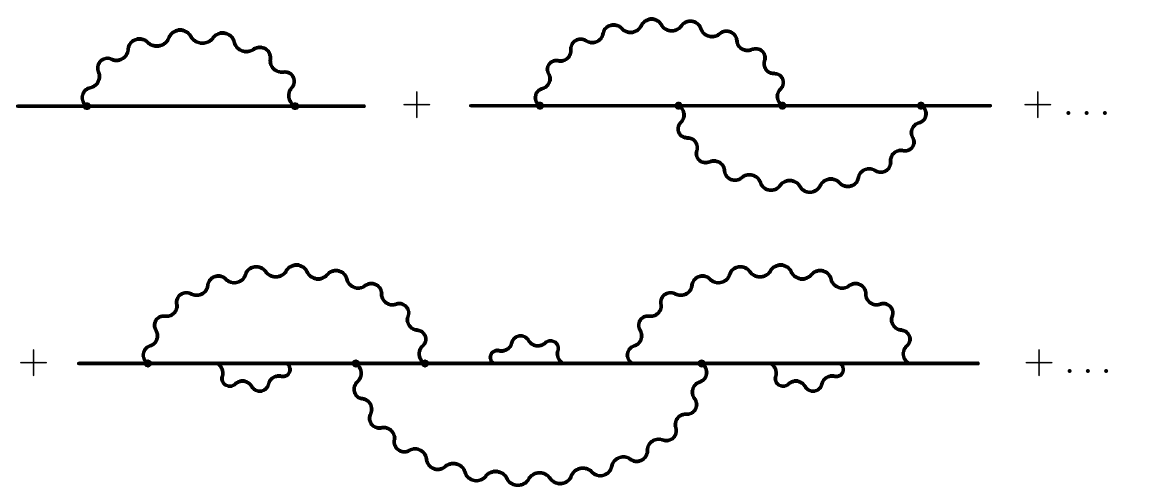}
\end{center}
\caption{Radiative corrections to the self-energy.\label{fig4}}
\end{figure}

In contrast to the two-body case, the system of equations for the
three-body truncation is rather non-trivial. For example, its
iteration generates all the graphs for the self-energy which
contain one fermion and two bosons, including overlapping
self-energy type diagrams. The number of such irreducible graphs
is infinite. Some of them are shown in Fig.~\ref{fig4}. The
solution of the system of coupled equations incorporates the sum
of these contributions to all orders.

Note that due to the covariance of our approach, we can identify
the contribution $\sim {\sla \omega}$ to the self-energy which
explicitly depends on the light front plane orientation. If not
regularized, the coefficient ${\cal C}(k^2)$ is quadratically
divergent and needs {\it a priori} both PV fermion and boson
regularization. After this regularization however, ${\cal
C}(k^2)\equiv 0$ for any values of the PV fermion and boson
masses. This makes the two-body self-energy identical to the
result obtained in perturbation theory in the Feynman
approach~\cite{kms2004}. The contributions ${\cal A}$ and ${\cal
B}$ do respect chiral symmetry in the sense that they are equal to
zero when the constituent mass $m$, as well as the physical mass
$M$ (which, in our case, coincides with $m$), goes to zero,
without the need of an extra PV boson. This is at variance with
the standard formulation of LFD where it is claimed that an
additional PV boson is needed, if the PV fermion mass is kept
finite~\cite{ch_09}.

The parameters $g_{02}$ and $\delta m_2$ are taken from the $N=2$
calculation~\cite{kms_08}. They are given by
\begin{subequations}
\label{N2}
\begin{eqnarray}
\label{g02}
g_{02}^2&=&\frac{g^2}{1-g^2 J_2},\\
\label{dm2}
\delta m_2&=&g^2\left[ {\cal A}(m^2) +{\cal
B}(m^2)\right],
\end{eqnarray}
\end{subequations}
where
\begin{equation}
J_2=-\frac{{\cal B}(m^2)}{m}-z_0
\end{equation}
with $z_0=2m\left[{\cal A}'(m^2)+{\cal B}'(m^2)\right]$. The norms of the one-
and two-body Fock sectors, entering the normalization
condition~(\ref{Inor}), are
\begin{subequations}
\label{norme2}
\begin{eqnarray}
I_1^{(2)}&=&1-g^2J_2, \\
I_2^{(2)}&=&g^2J_2.
\end{eqnarray}
\end{subequations}

In the two-body approximation mentioned above and discussed in
detail in Ref.~\cite{kms_08}, the two-body vertex function is
automatically independent of $\omega$ since ${\cal C}(k^2)\equiv
0$. Moreover, it is a constant (i.~e. it does not depend on the
momenta of the constituent). Due to all these, the
renormalization condition~(\ref{rencond2}) directly leads to the
relation~(\ref{g02}) between the bare and physical coupling
constants. In principle, nothing prevents $\Gamma_2$ to be
$\omega$-dependent, since it is an off-shell object, but this dependence must completely
disappear on the energy shell, i.e. for  $s_2=m^2$. It would be
indeed so if no Fock space truncation occurs. The latter, in
approximations higher than the two-body one (i.~e. for
$N=3,\,4,\ldots$), may cause some $\omega$-dependence of
$\Gamma_2$ even on the energy shell, which immediately makes the
general renormalization condition~(\ref{Gam3a}) ambiguous. If
so, one has to insert into the light-front interaction
Hamiltonian new counterterms which explicitly depend on
$\omega$ and cancels the $\omega$-dependence of
$\Gamma_2(s_2=m^2)$. Its explicit form will be given in the next
subsection. Note that the explicit covariance of CLFD allows to
separate the terms which depend on the light front plane
orientation (i.e. on $\omega$) from other contributions and
establish the structure of these counterterms. This is not
possible in ordinary LFD.

\subsection{Calculation of the two-body Fock component}
The method of solution is similar to that used in the
calculation~\cite{kms_08} for $N=2$. We first decompose the
vertex functions in invariant amplitudes. The vertex functions
on the l.-h.~s.'s of Eqs.~(\ref{eq245}), being matrices in the
spin indices, can be decomposed in a full set of spin matrices.
This decomposition is very simple in CLFD and takes the form
\begin{eqnarray}
\bar{u}(p_{1i})\Gamma_{1}^{i}u(p) & = &
(m_{i}^2-M^2)\psi_{1}^{i}\bar{u}(p_{1i})u(p),
\nonumber\\
\bar{u}(k_{1i})\Gamma_{2}^{ij}u(p) & = &
\bar{u}(k_{1i})\left[b_{1}^{ij}+ b_{2}^{ij} {\displaystyle
\frac{m{\sla \omega}}{\omega\cd p}}\right]u(p),
\nonumber\\
&& \label{gams}
\end{eqnarray}
where $\psi_1^i$ is a constant, and $b_{1,2}^{ij}$ are invariant
functions of particle momenta, with $m_0\equiv m$ and $\mu_0\equiv
\mu$. We denote temporarily, in the first of the above equations,
the physical fermion mass by $M$ in order to avoid singularities,
since the equations contain the combination
$\Gamma_1^i/(m_i^2-M^2)$ which becomes indeterminate for $i=0$ at
$M=m$. Using $M\neq m$ allows to take a smooth limit $\lim_{M\to
m}[\Gamma_1^i/(m_i^2-M^2)]=\psi_1^i$; after that one may set
$M=m$.

Each of the functions $b^{ij}_{1,2}$ depends on two invariant
kinematical variables. As usual, we define a pair of variables,
consisting of the longitudinal and transverse (with respect to the
three-vector ${\bg \omega}$ ) momenta:
\begin{equation} \label{lf}
x=\frac{\omega \cd k_2}{\omega \cd p} \ \ , \ \ {\bf
R}_{\perp}={\bf k}_{2\perp}-x {\bf p}_{\perp},
\end{equation}
where $k_2$ is the boson four-momentum. Then $b^{ij}_{1,2}$ are
functions of $x$ and $R_{\perp}^2$.

The renormalization condition~(\ref{Gam3c}), for $N=3$, implies
two conditions
\begin{subequations}
\label{b12on}
\begin{eqnarray}
\label{b12on1}
b_1^{00}(s_2=m^2)&=&g\sqrt{I_1^{(2)}},\\
\label{b12on2}
b_2^{00}(s_2=m^2)&=&0
\end{eqnarray}
\end{subequations}
for the spin components of $\Gamma_2$ at $s_2=m^2$, where the
two-body invariant energy squared $s_2$ is expressed through
$R_{\perp}$ and $x$ as follows:
\begin{equation}
\label{s2}
s_2=\frac{R_{\perp}^2+\mu^2}{x}+\frac{R_{\perp}^2+m^2}{1-x}.
\end{equation}
One should emphasize that the renormalization conditions are
imposed on the two-body vertex function $\Gamma_2^{00}$
corresponding to both physical constituents. The
condition~(\ref{b12on1}) defines unambiguously $g_{03}$. The
condition~(\ref{b12on2}) is  not verified automatically if the Fock
space is truncated for $N\ge 3$, unlike the case $N=2$. We
should thus enforce it by introducing an appropriate counterterm,
as explained above (see also Ref.~\cite{kms_08}). It corresponds
to the following additional structure in the interaction
Hamiltonian:
\begin{equation}
\label{Zomega} \delta {\cal H}^{int}_\omega = -Z_\omega \bar \psi'
\frac{m \sla \omega}{i \omega \cd
\partial} \psi'\varphi',
\end{equation}
where $Z_\omega$ is a constant adjusted to make Eq.~(\ref{b12on2})
true. The operator ${\sla \omega}/(i\omega\cd
\partial)$, in momentum space, leads to the appearance of a new
three-leg vertex ${\sla \omega}/(\omega\cd k)$ on each
fermion-boson vertex with
total incoming momentum $k$.
In principle, a
similar new $\omega$-dependent counterterm should be also added
to the Hamiltonian in order to cancel the $\omega$-dependence of
$\delta m_3$, in full analogy with the cancellation of
$b_2^{00}(s_2=m^2)$ \cite{kms2004}. However, as we will see
below, $\delta m_3$ is needed only for a calculation in the
four-body Fock space truncation. For this reason, working within
the three-body truncation only, we may not bother about
additional counterterms excepting that given by
Eq.~(\ref{Zomega}).

To solve the system of equations (\ref{eq245}), we substitute the decompositions~(\ref{gams}) into the expressions
for $V_{1-6}$ given in Appendix~\ref{V16}, then multiply
Eqs.~(\ref{eq24}) and~(\ref{eq25}) by $u(p_{1i})$ and $u(k_{1i})$,
respectively, to the left and each of them by $\bar{u}(p)$ to the
right, and sum over spin projections. We thus get
\begin{subequations}
\begin{widetext}
\begin{eqnarray}\label{eq24p}
(m_{i}^2-m^2)({\sla p}_{1i}+m_{i})\psi_{1}^{i}({\sla p}+m) & =
&({\sla p}_{1i}+m_{i})\left(V_{1}+V_{2}\right)({\sla p}+m),\\
({\sla k}_{1i}+m_{i})\left[b_{1}^{ij}+b_{2}^{ij}\frac{m{\sla
\omega}}{\omega\cdot p}\right]({\sla p}+m) &= &({\sla
k}_{1i}+m_{i})\left(V_{3}+V_{45}+V_{6}\right)({\sla p}+m).
\label{eq25p}
\end{eqnarray}
\end{widetext}
\end{subequations}
The system of matrix equations~(\ref{eq24p}) and~(\ref{eq25p}) can
be transformed into a homogeneous system of ten linear integral
equations for ten unknown functions (two $\psi_{1}^{i}$, four
$b_{1}^{ij}$, and four $b_{2}^{ij}$). These equations are obtained
by taking the trace of Eqs.~(\ref{eq24p}) and~(\ref{eq25p}) (six
equations), and by taking the trace of Eq.~(\ref{eq25p}) after the
multiplication of its both sides by ${\sla \omega}$ (four
equations).

In order to achieve the limit $m_1 \to \infty$, it is convenient
to  replace the functions $\psi_1^i$ and $b_{1,2}^{ij}$ by
the new functions $\alpha_i$, $h_i^j$, and $H_i^j$ according to
\begin{eqnarray}\label{r1}
\psi_1^i&=&\frac{m}{m_i(m+m_i)}\alpha_i,\quad b_1^{ij}=
\frac{m_i}{m}h_i^j,
\nonumber\\
b_2^{ij}&=&\frac{m_i}{m}
\frac{H_i^j-\left(1-x+\frac{m_i}{m}\right) h_i^j}{2(1-x)} \ .
\end{eqnarray}
A careful analysis shows that in this limit the PV mass $m_1$
disappears from the equations written in terms of $\alpha_i$,
$h_i^j$, and $H_i^j$. These functions have therefore a finite
limit. Below we will imply that the limit $m_1\to\infty$ is taken
and $\alpha_i$, $h_i^j$, and $H_i^j$ denote the limiting values.

For further simplification of the equations,
it is convenient to introduce new functions $\tilde{h}_{0,1}^j$
and $\tilde{H}_{0,1}^j$ by means of the relation
\begin{equation}
\label{func}
\left(
\begin{array}{c}
h_{0,1}^j\\
H_{0,1}^j
\end{array}
\right)=\alpha_0\kappa
\left(
\begin{array}{c}
\tilde{h}_{0,1}^j\\
\tilde{H}_{0,1}^j
\end{array}
\right)\ ,
\end{equation}
with
\begin{equation}
\label{kappa} \kappa=g_{03}\frac{
1-g^2J_2}{1+g^2z_0}.
\end{equation}
Using the substitution~(\ref{func}) and denoting
\begin{equation}
\label{Z3}
Z'_\omega=\frac{2Z_\omega}{g_{03}}-\frac{\alpha_1}{\alpha_0},
\end{equation}
the initial system of ten equations splits into two
subsystems. The first one contains two equations involving the
ratio $\alpha_1/\alpha_0$. The value of $\delta m_3$ just ensures
that both equations define the same quantity
$\alpha_1/\alpha_0$. It is not interesting for our study in the
three-body approximation, since $\alpha_1$, as will be seen below,
drops out from the observables we calculate here, while $\alpha_0$
is uniquely determined by the normalization condition for the
state vector. As already mentioned, we also do not need to calculate $\delta m_3$
itself.  It
is used as an input in the calculation at the next, $N=4$, truncation.

The second
subsystem of eight equations involves the eight functions
$\tilde{h}_i^j$ and $\tilde{H}_i^j$ only, since the ratio
$\alpha_1/\alpha_0$ is absorbed into the definition of $Z'_\omega$
in Eq.~(\ref{Z3}). We thus get
\begin{eqnarray}
\tilde{h}_0^j & = & 1+{g'}^2\left(K_1^j h_0^j+ K_2^j
\tilde{h}_1^j\right)+{g'}^2 i_0^j,
\nonumber\\
\tilde{h}_1^j & = & {g'}^2\left(-K_3^j \tilde{h}_0^j+ K_4^j
\tilde{h}_1^j\right)+{g'}^2 i_1^j,
\nonumber\\
\tilde{H}_0^j & = & Z'_\omega(1-x)+2-x
\label{hH} \\
&+&{g'}^2\left(K_1^j \tilde{H}_0^j+ K_2^j
\tilde{H}_1^j\right)+{g'}^2 I_0^j,
\nonumber\\
\tilde{H}_1^j & = & 1+{g'}^2\left(-K_3^j \tilde{H}_0^j+ K_4^j
\tilde{H}_1^j\right)+{g'}^2 I_1^j, \nonumber
\end{eqnarray}
where
\begin{equation} \label{gp}
{g'}^2=\frac{g^2}{(1+g^2z_0)},
\end{equation}
and
\begin{eqnarray*}
K_1^j & = & \frac{1}{m}\left\{{\cal
B}_r(s_1)-\frac{2[{\cal A}_r(s_1)+{\cal B}_r(s_1)]m^2}
{m^2-s_1}\right\},\\
\nonumber\\
K_2^j & = & \frac{{\cal A}_r(s_1)+{\cal B}_r(s_1)}{m},\\
\nonumber\\
K_3^j & = &
\frac{[{\cal A}_r(s_1)+{\cal B}_r(s_1)]m}
{m^2-s_1},\\
\nonumber\\
K_4^j & = & \frac{{\cal B}_r(s_1)}{m}.\\
\nonumber
\end{eqnarray*}
The 
substracted self-energy contributions 
${\cal
A}_r(s_1)$ and ${\cal B}_r(s_1)$,  are given by
$$
{\cal A}_r(s_1)={\cal A}(s_1)-{\cal A}(m^2),\,\,\,\,\,\,\,\, {\cal
B}_r(s_1)={\cal B}(s_1)-{\cal B}(m^2)
$$
with
\begin{equation}
s_1=-\frac{{R}_{\perp}^2}{x}+(1-x)m^2-\frac{1-x}{x}\mu_j^2.
\end{equation}
The
functions ${\cal A}$ and ${\cal B}$ are given
in Appendix~\ref{secoef}, while
the integral terms $i_{0,1}^j$ and
$I_{0,1}^j$ are  given in Appendix~\ref{app1}. 

The limit of infinite PV mass $\mu_1$ is not easy to perform
analytically, as it was done for $m_1$. Setting $\mu_1\to\infty$
directly  in Eqs.~(\ref{hH}) makes some integration kernels
singular (they decrease too slowly at $R_{\perp}\to\infty$). The
dependence of
physical observables, like the
AMM, on $\mu_1$ will therefore be studied numerically.

Note that although $g_{02}^2$ in Eq.~(\ref{g02}) can become
infinite (for $J_2=1/g^2$) and changes sign from positive to
negative at  sufficiently large values of the PV boson mass
$\mu_1$, the eigenvalue equations~(\ref{hH}) do not show any
singularity when $g_{02}^2$ goes to infinity. Indeed, $g_{02}^2$
does not appear in the equations~(\ref{hH}). These equations
depend only on ${g'}^2$, given by Eq.~(\ref{gp}), with
$z_0$ strictly positive. Therefore, $g'^2$ is strictly positive and
finite. We shall come back in Sec.~\ref{num} to the interpretation
of the limit of large $\mu_1$, when both $g_{02}^2$ and the norm of the
one-body sector $I_1^{(2)}$ are negative, while the norm of the two-body
sector $I_2^{(2)}$ is larger than $1$, from Eqs.~(\ref{norme2}).

The constant
$Z'_{\omega}$ entering the system of
equations~(\ref{hH}) is determined from the renormalization
condition~(\ref{b12on2}), while $g_{03}$ needed to calculate the renormalized 
vertex functions in Eqs.~(\ref{func}) is determined from the renormalization condition~(\ref{b12on1}).

The components $b_{1,2}^{00}(s_2=m^2)$ entering these
renormalization  conditions are expressed through the
solution of the system of equations~(\ref{hH}) by means of
Eqs.~(\ref{r1}) and~(\ref{func}). The kinematical point $s_2=m^2$
belongs to a nonphysical region, but there is no need to make an
analytical continuation to this region of the solution
$\tilde{h}_i^j$ and $\tilde{H}_i^j$ found numerically. Indeed, the
integral terms in Eqs.~(\ref{hH}) involve integrations within the
physical domain only. One can simply set $j=0$,
$R_{\perp}=R_{\perp}^*$, $x=x^*$, where $R_{\perp}^*$ and $x^*$
are determined by the condition $s_2=m^2$, and calculate the
integral terms by substituting there the previously found solution
$\tilde{h}_i^j(R_{\perp},x)$ and $\tilde{H}_i^j(R_{\perp},x)$ for
physical values $R_{\perp}$ and $x$. After that Eqs.~(\ref{hH})
reduces to a system of four ordinary linear inhomogeneous
equations for $\tilde{h}_i^0(R_{\perp}^*,x^*)$ and
$\tilde{H}_i^0(R_{\perp}^*,x^*)$. Finally, relating the calculated
quantities $\tilde{h}_0^0(R_{\perp}^*,x^*)$ and
$\tilde{H}_0^0(R_{\perp}^*,x^*)$ to $b_{1,2}^{00}(s_2=m^2)$ we
get $g_{03}$ from Eq.~(\ref{b12on1})\footnote{More precisely, we
get not $g_{03}$ alone but the product $g_{03}\alpha_0$. The
quantity $\alpha_0$ is found from the normalization condition for
the state vector. This procedure requires knowing the three-body
normalization integral which is calculated in the next section.}
and $Z'_{\omega}$ from Eq.~(\ref{b12on2}).

The condition $s_2=m^2$ however does not determine $R_{\perp}^*$
and $x^*$ simultaneously. It is convenient to fix $x^*$ somehow
and then find $R_{\perp}^*$
from Eq.~(\ref{s2}):
\begin{equation}
\label{rstar} R^{*2}_\perp=-\left[x^{*2} m^2+(1-x^*) \mu^2 \right].
\end{equation}
Since the two-body vertex
function~(\ref{gams}) on the energy shell must turn into a
constant, the functions $b_{1,2}^{00}(s_2=m^2)$ also must be
constants. In other words, if one relates the arguments of these
functions by Eq.~(\ref{rstar}), their values are independent of
the choice of $x^*$. It would be so in exact calculations, i.~e.
if Fock space was not truncated. A finite order truncation makes the
Fock components, even at $s_2=m^2$, $x^*$-dependent. As advocated in
Ref.~\cite{kms_08}, we choose $x^*=\frac{\mu}{m+\mu}$. We shall
investigate in Sec.~\ref{num} how $b_{1,2}^{00}(s_2=m^2)$ depends
on the choice of $x^*$.

\subsection{Representation of the three-body component}
We can find the three-body
component by calculating the amplitude corresponding to the r.-h.~s. of
the equation shown by the last line  in Fig.~\ref{syst_eq}.

The general form of the relativistic vertex function of a system
composed from one constituent fermion and two spinless bosons with total
spin 1/2 reads
\begin{equation}
\label{wf1}
\bar{u}^{\alpha}_{\sigma_1}(k_{1}) \Gamma_{\alpha\beta}(1,2,3)
u_{\sigma}^{\beta}(p),
\end{equation}
where $\Gamma_{\alpha\beta}(1,2,3)$ is a $4\times 4$-matrix in the
indices $\alpha,\beta$. The arguments of $\Gamma_3$, denoted
symbolically by numbers, mean three pairs of the standard
variables
$$
{\bf R}_{l\perp}={\bf k}_{l\perp}-x_l{\bf p}_{\perp},\,\,\,\,\,\,\,\,
x_l=\frac{\omega\cd k_l}{\omega\cd p}
$$
with $l=1$ corresponding to the fermion and $l=2,\,3$ to
bosons. Here $\bar{u}^{\alpha}_{\sigma_1}(k_{1})$ is the bispinor
of the constituent fermion, $u_{\sigma}^{\beta}(p)$ is the
bispinor of the physical fermion (of the composite system),
$\sigma_1$, $\sigma$ are their spin projections in the
corresponding rest frame. Since $\sigma_1=\pm 1/2$ and $\sigma=\pm
1/2$, we have in general $2\times 2=4$ matrix elements. Usually,
parity conservation reduces  this number by a factor of two. However, this is
not the case in relativistic calculations, for a $n$-body wave
function with $n \ge 3$~\cite{karm98}. This wave function is
determined by {\em four independent} matrix elements or,
equivalently, by four scalar functions $g_{1-4}$ like
\begin{eqnarray}
\label{wf2}
&&\bar{u}(k_{1})\Gamma_3(1,2,3) u(p) \\
 &&=\bar{u}(k_{1})\Bigl(g_1\,S_1+g_2\,S_2+g_3\,S_3+g_4\,S_4\Bigr)
u(p).\nonumber
\end{eqnarray}
For simplicity, we omitted the bispinor indices and the indices
marking the particle type (either physical or PV one). It is
convenient to construct the four basis spin structures as follows:
\begin{eqnarray}
\label{base1}
S_1&=&2x_1-(m_{i}+x_1 m) \displaystyle{\frac{\sla
\omega}{\omega\cd p}},
\nonumber \\
S_2&=&m\displaystyle{\frac{\sla \omega}{\omega\cd p}},
\nonumber\\
S_3&=&iC_{ps}\left[2x_1 -(m_{i}-x_1 m )\displaystyle{\frac{\sla
\omega}{\omega\cd p}}\right]\gamma_5,
\nonumber\\
S_4&=&i\,m\,C_{ps}\displaystyle{\frac{\sla \omega}{\omega\cd
p}}\gamma_5
\end{eqnarray}
with $x_1=\frac{\omega \cd k_1}{\omega \cd p}$, and $m_{i}$ being
the internal fermion mass (either physical or PV one, depending on
which type of fermion the momentum $k_1$ corresponds to), while
$C_{ps}$ is the following pseudoscalar:
\begin{equation}
\label{wf6}
C_{ps}=\frac{1}{m^2\omega\cd p}
e^{\mu\nu\rho\gamma}k_{2\mu}k_{3\nu}p_{\rho}\omega_{\gamma}.
\end{equation}
The
function $C_{ps}$ can only be constructed with
four independent four-vectors. This is the case in LFD for $n \ge
3$. In the non-relativistic limit, one would need $n \ge 4$. We
can then construct two additional spin structures $S_3$ and $S_4$
of the same parity as $S_1$ and $S_2$ by combining $C_{ps}$  with
parity negative matrices constructed from $S_1$, $S_2$, and
$\gamma_5$ matrices. Instead of $k_{2\mu}k_{3\nu}$ one could have
taken any pair of momenta ($k_{1\mu}k_{3\nu}$ or
$k_{1\mu}k_{2\nu}$). We take the boson momenta for symmetry. With
this definition
\begin{equation}
\label{Cps2} C_{ps}^2 =\frac{1}{m^4}[{ R}^2_{2\perp}{
R}^2_{3\perp}-({\bf R}_{2\perp}\cd {\bf R}_{3\perp})^2].
\end{equation}
The three-body vertex function $\Gamma_3(1,2,3)$ is completely
determined by the four scalar functions $g_{1-4}(1,2,3)$ in
Eq.~(\ref{wf2}). They depend on ${\bf R}_{1-3,\perp}$ in the form
of their scalar products among themselves and on $x_{1-3}$. Since
\begin{equation}
\label{Rx}
{\bf R}_{1\perp}+{\bf R}_{2\perp}+{\bf R}_{3\perp}=0,\quad
x_1+x_2+x_3=1,
\end{equation}
we can exclude, for instance,
${\bf R}_{1\perp}$ and $x_1$.
The functions $g_{1,2}$ are symmetric relative to the
permutation $2\leftrightarrow 3$, whereas $g_{3,4}$ are
antisymmetric:
\begin{eqnarray*}
g_{1,2}(1,2,3)&=&g_{1,2}(1,3,2), \\
g_{3,4}(1,2,3)&=-&g_{3,4}(1,3,2),
\end{eqnarray*}
so that the product $C_{ps}\ g_{3,4}(1,2,3)$ which appears in
$S_{3}\ g_{3}$ and $S_{4}\ g_{4}$ is symmetric.

Each component $g_n$ is represented as a sum or a difference of
two terms:
\begin{eqnarray}
\label{gn1}
&&g_{1,2}(1,2,3)=\bar{g}_{1,2}(1,2,3)+\bar{g}_{1,2}(1,3,2),
\nonumber\\
&&g_{3,4}(1,2,3)=\bar{g}_{3,4}(1,2,3)-\bar{g}_{3,4}(1,3,2),
\end{eqnarray}
where the permutation $2\leftrightarrow 3$ means
$${\bf R}_{2\perp}\leftrightarrow {\bf R}_{3\perp},\quad
x_2 \leftrightarrow x_3,\quad \mu_{j_2} \leftrightarrow\mu_{j_3}.
$$
In their turn, $\bar{g}_n(1,2,3)$, according to the last line in
Fig.~\ref{syst_eq}, are linearly expressed through the functions
$\tilde{h}^{j_2}_{0,1}$ and $\tilde{H}^{j_2}_{0,1}$ which form a
solution of the equations (\ref{hH}):
\begin{multline}
\label{gn2}
\bar{g}_n(1,2,3) =
\alpha_0\ \kappa \ g_{02}\left[a_{n0}(1,2,3) \tilde{ h}^{j_2}_0(2)
\right. \\
\left.+a_{n1}(1,2,3)  \tilde{h}^{j_2}_1(2)
+
A_{n0}(1,2,3)  \tilde{H}^{j_2}_0(2) \right.\\ + \left.
A_{n1}(1,2,3)
\tilde{H}^{j_2}_1(2)\right].
\end{multline}
The coefficients $a$ and $A$ in this formula are given in
Appendix~\ref{app4}.

We can finally calculate the three-body normalization integral. It is given by
\begin{eqnarray}
\label{I3}
I_3&=&\frac{1}{2}\sum_{j_2,j_3 =0}^{1}(-1)^{j_2+j_3}\int
\frac{n_3^{j_2 j_3}} {(s_3-m^2)^2}\; dD_3
\end{eqnarray}
with
\begin{eqnarray}\label{norm1}
n_3^{ j_2 j_3} &=&\frac{1}{2} \mbox{Tr}[\bar{\Gamma}_3({\sla
k}_1+m)\Gamma_3({\sla p}+m)]\\
&=& 4x_1[R_{1\perp}^2 g_1^2+m^2g_2^2+C_{ps}^2(R_{1\perp}^2 g_3^2
+m^2 g_4^2)],\nonumber
\end{eqnarray}
and, as usual, $\bar \Gamma = \gamma^0 \Gamma ^\dagger \gamma^0$.
The factor $\frac{1}{2}$ in Eq.~(\ref{norm1}) results from
averaging over initial state spin projections, while the factor
$\frac{1}{2}$ in Eq.~(\ref{I3}) is the combinatorial factor
$\frac{1}{(n-1)!}$ originating from the identity of the two
bosons. The contribution of PV fermion is omitted since it
disappears in the limit $m_1\to\infty$. The phase space volume
element has the form (see Eq.~(3.23) from Ref.~\cite{cdkm}):
\begin{eqnarray}
\label{D3}
dD_3&=&2(2\pi)^3\delta^{(2)}({\bf R}_{1\perp}+ {\bf R}_{2\perp}+
{\bf R}_{3\perp})\\
&\times&\delta(x_1+x_2+x_3-1)\prod_{l=1}^3
\frac{d^2R_{l\perp}dx_l}{(2\pi)^3 2x_l}. \nonumber
\end{eqnarray}
%

\section{Electromagnetic form factors}  \label{elm}
\subsection{Electromagnetic vertex in CLFD}
\begin{figure}[ht!]
\begin{center}
\includegraphics[width=8.5cm]{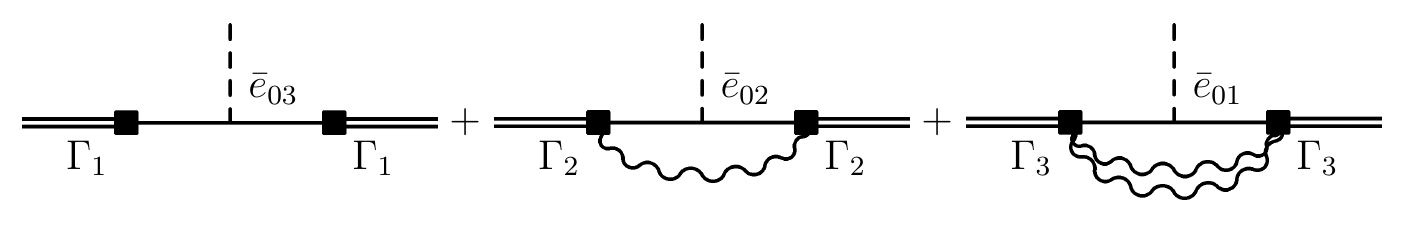}
\end{center}
\caption{One-, two-, and three-body contributions to the
electromagnetic vertex.\label{em_v}}
\end{figure}

The electromagnetic vertex contains contributions of \mbox{one-,}
two-, and three-body Fock sectors, as shown in Fig.~\ref{em_v}.
They are expressed, in our FSDR scheme, in terms of the external
electromagnetic BCC $\bar e_{0l}$, as explained in
Ref.~\cite{kms_08}. These coupling constants are all identical to
the physical
fermion charge, i.~e., $\bar{e}_{0l}=e$ for all
$l$'s. Note that this important property of QED is not preserved
in general if FSDR is not used.

The decomposition of the spin-1/2 electromagnetic vertex  in CLFD
is given by~\cite{km96, kms2007}
\begin{widetext}
\begin{equation}
\label{F12} \bar{u}(p')G^{\rho}u(p)=
e\bar{u}(p')\left[F_1\gamma^{\rho}+
\frac{iF_2}{2m}\sigma^{\rho\nu}q_{\nu} +B_1\left( \frac{{\sla
\omega}}{\omega\cd
p}P^{\rho}-2\gamma^{\rho}\right)+B_2\frac{m{\omega^{\rho}}}{\omega\cd
p}+B_3\frac{m^2{\sla \omega}\omega^{\rho}}{(\omega\cd
p)^2}\right]u(p)
\end{equation}
\end{widetext}
with $P=p+p'$,  and $q=p'-p$. $F_1$ and $F_2$ are the physical
form factors, while $B_{1,2,3}$ are spurious (nonphysical)
contributions which appear if rotational invariance is broken,
e.~g. by Fock space truncation. The decomposition~(\ref{F12})
enables to separate unambiguously the physical form factors from
the nonphysical ones. Under the condition $\omega\cd q=0$, all
$F_{1,2}$, $B_{1-3}$ depend on $Q^2\equiv -q^2$ only.

We shall represent $q=(q_0, {\bf \Delta}, q_\parallel)$, where
$q_\parallel$ and ${\bf \Delta}$ are, respectively, the
longitudinal and transverse components of the momentum transfer
with respect to the three-vector ${\bf
\omega}$. Since $\omega\cd q=\omega_0(q_0-q_\parallel)=0$, we have
$Q^2={\bf \Delta}^2$.

After construction of the matrix
\begin{equation}
\label{O} O^{\rho}=\frac{1}{4m^2}({\sla p}'+m)G^{\rho}({\sla
p}+m),
\end{equation}
and calculation of the traces
\begin{equation}
\label{ff20} c_4=\mbox{Tr}[O_{\rho}\omega^{\rho}]m/\omega\cd
p,\quad c_5=\mbox{Tr}[O_{\rho}\omega^{\rho}{\sla
\omega}]m^2/(\omega\cd p)^2,
\end{equation}
the electromagnetic form factors write~\cite{kms2004}
\begin{equation}
\label{F1F2}
F_1=\frac{1}{2}c_5,\quad F_2=\frac{2m^2}{Q^2}(c_5-c_4).
\end{equation}
The value $F_1(Q^2=0)$ equals one, since it coincides with the norm of the
state vector. The value $F_2(Q^2=0)$ is just the
AMM.

\subsection{One-body contribution}
The one-body contribution to the form factor $F_1$ is given by the
first diagram in Fig.~\ref{em_v}. It does not depend on $Q^2$ and
coincides with the norm of the one-body sector:
\begin{equation}
\label{F11b}
F_{1,1b}=\alpha_0^2 \equiv I_1.
\end{equation}
There is no one-body contribution to the form factor $F_2$:
\begin{equation}
\label{F21b}
F_{2,1b}=0.
\end{equation}
%

\subsection{Two-body contribution}
The two-body contribution to the electromagnetic vertex, as given
by the second diagram in Fig.~\ref{em_v}, writes
\begin{multline*}
\bar u(p')
 G^{\rho}_{2b}
 u(p) =
\frac{1}{(2\pi)^3} \sum_{i,i',j=0}^{1} (-1)^{i+i'+j}
\\
\times\int d^2R_{\perp}\int_0^1 \frac{dx}{2x(1-x)^2}
\\
\times \frac{\bar u(p') \bar \Gamma^{'{i'j}}_2({\sla k}'_{1
i'}+m_{i'})\gamma^{\rho} ({\sla k}_{1i}+m_i)\Gamma^{ij}_2 u(p)}
{({s'}^{i'j}_2-m_{i'}^2) ({s}^{ij}_2-m_{i}^2)},
\end{multline*}
where $\Gamma^{ij}_2$ is given by Eq.~(\ref{gams}) with
$b_{1,2}^{ij}=b_{1,2}^{ij}(R_{\perp}^2,x)$, $k_{1i}$ ($k'_{1i'}$)
is the momentum of the constituent fermion incoming to (outgoing
from) the elementary electromagnetic vertex,
$s_2^{ij}=(k_{1i}+k_{2j})^2$, $s_2^{ \prime
i'j}=(k_{1i'}'+k_{2j})^2$, and $k_{2j}$ is the constituent boson
momentum. $\Gamma^{'{i'j}}_2$ has the same decomposition as
$\Gamma^{ij}_2$, with the replacement
$b^{ij}_{1,2}(R_{\perp}^2,x)\to b^{i'j}_{1,2}({R'}_{\perp}^2,x)$
with ${\bf R}'_{\perp}={\bf R}_{\perp}-x{\bf \Delta}$.

Using the relations~(\ref{r1}) and (\ref{func}), we can calculate
the two-body contribution to the electromagnetic form factors in
terms of the solutions $\tilde h_i^{j}$, $\tilde H_i^{j}$ of the
system of eigenvalue equations~(\ref{hH}). After taking the limit
$m_1\to\infty$ the result is as follows:
$$
F_{1,2b}=\frac{\alpha_0^2\kappa^2}{16\pi^3}\sum_{j=0}^1
(-1)^j\int d^2R_{\perp}\int_0^1 dx
$$
\begin{equation}
\label{F12b} \times\frac{x[({\bf R}_{\perp}\cd{\bf
R}'_{\perp})\tilde{h}_0^j \tilde{h}_0^{'j}+m^2\tilde{H}_0^j
\tilde{H}_0^{'j}]}
{[R_{\perp}^2+x^2m^2+(1-x)\mu_j^2][{R'}_{\perp}^2+x^2m^2+(1-x)\mu_j^2]},
\end{equation}
$$
F_{2,2b}=\frac{\alpha_0^2\kappa^2m^2}{4\pi^3\Delta^2}\sum_{j=0}^1
(-1)^j\int d^2R_{\perp}\int_0^1 dx
$$
\begin{equation}
\label{F22b} \times\frac{x({\bf R}_{\perp}\cd{\bf
\Delta})\tilde{h}_0^j \tilde{H}_0^{'j}}
{[R_{\perp}^2+x^2m^2+(1-x)\mu_j^2][{R'}_{\perp}^2+x^2m^2+(1-x)\mu_j^2]}.
\end{equation}
Functions with primes depend on ${R'}_{\perp}^2$ and $x$. The
value $F_{1,2b}(Q^2=0)$ coincides with the two-body contribution
to the normalization integral,
$$
I_2=\frac{\alpha_0^2\kappa^2}{16\pi^3}\sum_{j=0}^1
(-1)^j\int_0^{\infty} d^2R_{\perp}\int_0^1 dx\,x
$$
\begin{equation}
\label{F12b0}
\times\frac{R_{\perp}^2\left(\tilde{h}_0^j\right)^2
+m^2\left(\tilde{H}_0^j\right)^2}
{[R_{\perp}^2+x^2m^2+(1-x)\mu_j^2]^2}.
\end{equation}

To calculate the two-body contribution to the AMM, which is given
by the value $F_{2,2b}(Q^2=0)$, one should go over to the limit
$\Delta\to 0$ in Eq.~(\ref{F22b}). The corresponding analytic
formula includes derivatives over $R_{\perp}$ from the Fock
components. For numerical calculations it is however more
convenient to find $F_{2,2b}$ at small but finite $Q^2$ and then,
decreasing the latter, to reach desired accuracy.
The result of this numerical limiting
procedure is very stable.

\subsection{Three-body contribution}
The three-body contribution to the electromagnetic vertex reads
\begin{multline}\label{emv}
\bar u(p')
G^{\rho}_{3b}u(p)=\frac{1}{2}\sum_{j_2,j_3=0}^{1}(-1)^{j_2+j_3}
\\
\times \int\frac{\bar u(p') \bar{\Gamma'}_3({\sla k}'_{1}+m)
\gamma^{\rho}({\sla k}_{1}+m)\Gamma_3 u(p)}{4x_1^2x_2x_3(s_3-m^2)
(s'_3-m^2)}{d}D_3,\nonumber
\end{multline}
where $dD_3$ is defined by Eq.~(\ref{D3}) and
\begin{equation}
\label{s3}
s_3=\frac{{ R}_{1\perp}^2+m^2}{x_1}+\frac{{
R}_{2\perp}^2+\mu_{j_2}^2}{x_2} +\frac{{
R}_{3\perp}^2+\mu_{j_3}^2}{x_3}.
\end{equation}
$s'_3$ differs from $s_3$ by the following shift of the arguments:
\begin{eqnarray}
\label{Rp}
{\bf R}_{1\perp}&\to& {\bf R'}_{1\perp}={\bf
R}_{1\perp}+(1-x_1){\bf \Delta},
\nonumber\\
{\bf R}_{2\perp}&\to&{\bf R'}_{2\perp}={\bf R}_{2\perp}-x_2{\bf
\Delta},
\nonumber\\
{\bf R}_{3\perp}&\to&{\bf R'}_{3\perp}={\bf R}_{3\perp}-x_3{\bf
\Delta}.
\end{eqnarray}
$\bar{\Gamma'}_3$ is obtained from $\Gamma_3$ by the same shift of
arguments. From the
decomposition~(\ref{wf2}), we can calculate $G^{\rho}$, and
construct the matrix $O^{\rho}$ by means of Eq.~(\ref{O}). The
form factors are thus given by Eqs.~(\ref{F1F2}) and read
\begin{subequations}
\label{FF12}
\begin{eqnarray}
\label{F1}
F_{1,3b}&=& \int (C^{(1)}_{11}g_1g'_1 + C^{(1)}_{22}g_2g'_2 +
C^{(1)}_{33}g_3g'_3 \nonumber\\
&+& C^{(1)}_{44}g_4g'_4 +2C^{(1)}_{31}g_3 g'_1 )\frac{dD_3}{d_1},
\\
\label{F2}
F_{2,3b}&=& 2\int (C^{(2)}_{12}g_1g'_2 +C^{(2)}_{41}g_4g'_1 +
C^{(2)}_{32}g_3g'_2
\nonumber\\
&+&C^{(2)}_{34}g_3g'_4)\frac{dD_3}{d_2},
\end{eqnarray}
\end{subequations}
where
$$
d_1=m^4x_2 x_3(m^2-s_3)(m^2-s'_3), \quad d_2=\frac{2{
\Delta}^2}{m^2}d_1.
$$
$g'_n$ differs from $g_n$ by the shift of the arguments~(\ref{Rp}).
The coefficients $C_{nk}^{(1,2)}$ in Eqs.~(\ref{F1})
and~(\ref{F2}) are given in Appendix~\ref{app3}.

The value $F_{1,3b}(Q^2=0)$ coincides with the
norm, $I_3$, of  the three-body sector given by Eq.~(\ref{I3}).
The quantity $\alpha_0$ which has been unknown up to now, is determined
from the normalization condition for the state vector:
\begin{equation}
\label{alpha0}
\alpha_0^2+I_2+I_3=1.
\end{equation}
Since both $I_2$ and $I_3$ are proportional to $\alpha_0^2$, then,
denoting $I_{2,3}\equiv
\alpha_0^2\kappa^2\tilde{I}_{2,3}$,
where $\kappa$ is defined by Eq.~(\ref{kappa}),
we immediately get
\begin{equation}
\label{alpha0a}
\alpha_0^2=\frac{1}{1+\kappa^2(\tilde{I}_2+\tilde{I}_3
)}.
\end{equation}
%

\section{Numerical results}  \label{num}
The solution of Eqs.~(\ref{hH}) is found by a
matrix inversion after discretization of
the integrals, using Gaussian method. All integrals are finite at
finite PV boson mass $\mu_1$. As already mentioned, the limit of
infinite PV fermion mass $m_1$ has been done analytically, while
the Fock components $\tilde h_i^{j}$, $\tilde H_i^{j}$ and, hence,
$b_{1,2}^{ij}$ in Eq.~(\ref{gams}) do depend on the PV boson mass
$\mu_1$. The numerical calculations have been performed on an
ordinary modern laptop.

The AMM is calculated for a typical set of physical parameters
$m=0.938$ GeV, $\mu = 0.138$ GeV, and two values of the coupling
constant $\alpha \equiv \frac{g^2}{4\pi}=0.2$ and $0.5$. This
mimics, to some extent, a physical nucleon coupled to scalar
"pions". The typical pion-nucleon coupling constant is given by
$g=\frac{g_A}{2F_\pi}\langle k\rangle$ where $\langle k\rangle$ is
a typical momentum scale, and $g_A$ and $F_\pi$ are the axial
coupling constant and the pion decay constant, respectively. For
$\langle k\rangle = 0.2$ GeV we just get $\alpha \simeq 0.2$.
\begin{figure}[ht!]
\begin{center}
\includegraphics[width=8.5cm]{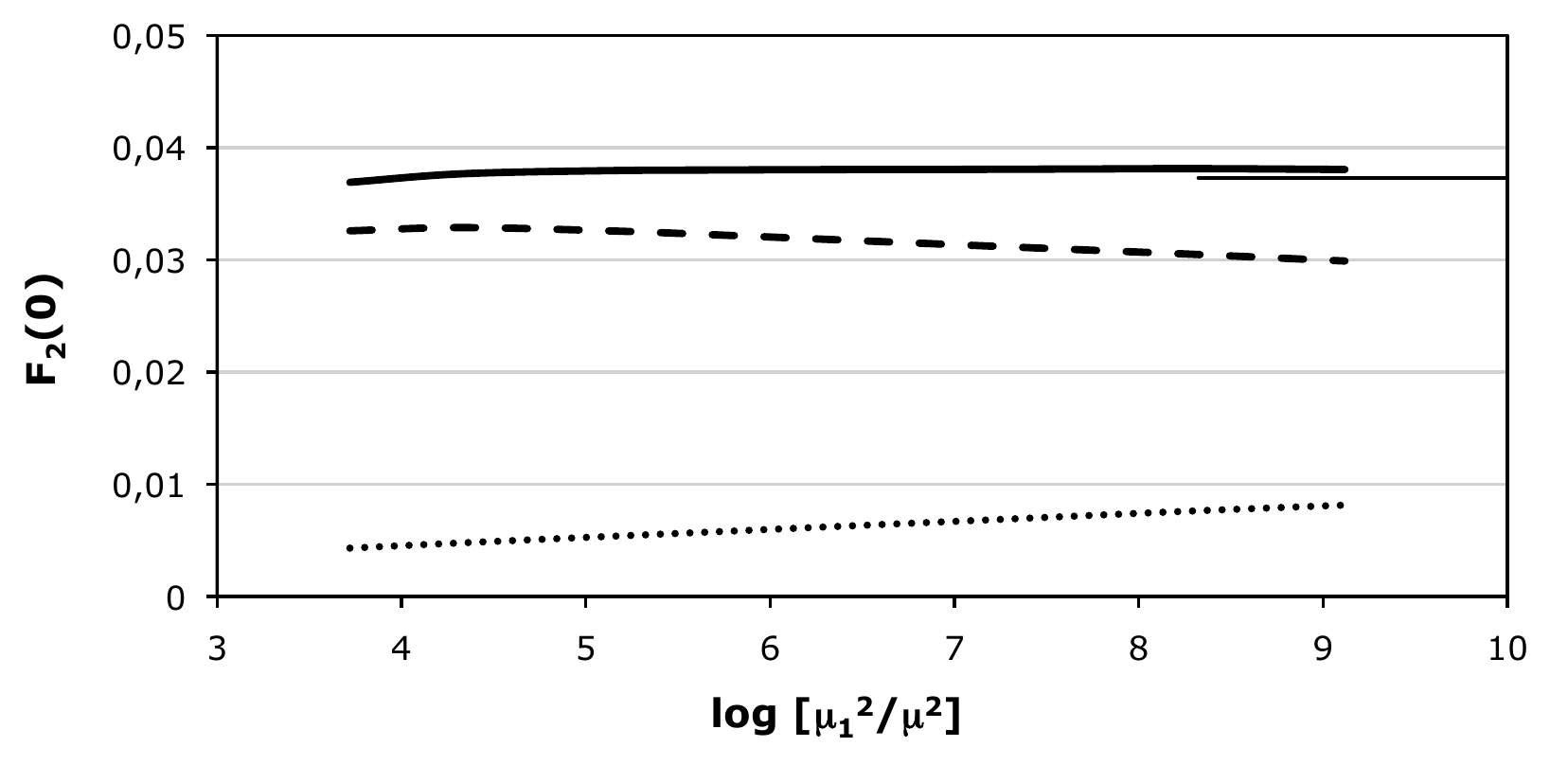}
\includegraphics[width=8.5cm]{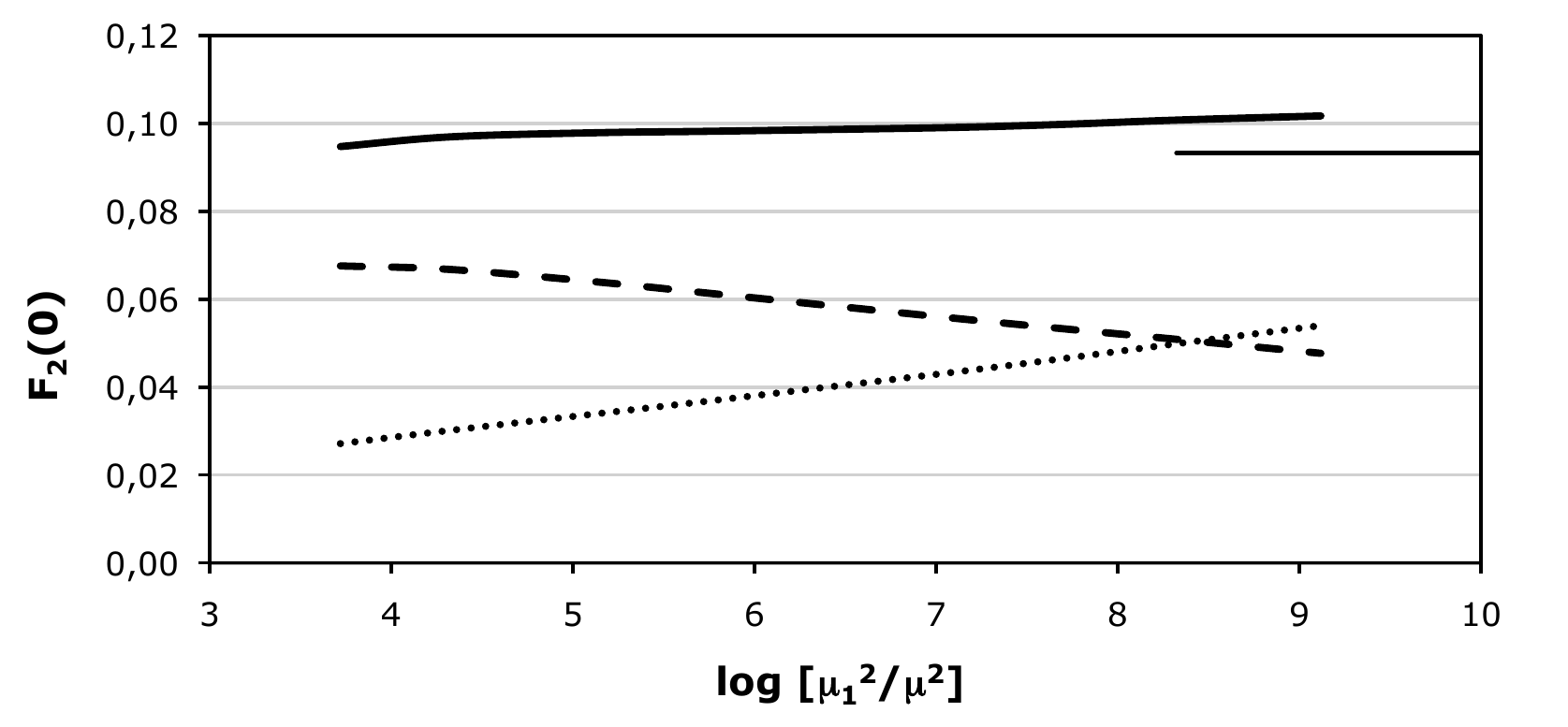}
\end{center}
\caption{The anomalous magnetic moment in the Yukawa model as a
function of the PV mass $\mu_1$, for two different values of the
coupling constant, $\alpha = 0.2$ (upper plot)
and $0.5$ (lower plot).
The dashed and dotted lines are, respectively, the two- and
three-body contributions, while the solid line is the total
result. The AMM value
calculated in the $N=2$ approximation is shown by the thin line on
the right axis.}\label{amm}
\end{figure}

We plot in
Fig.~\ref{amm} the AMM as a function of
$\log\left[\frac{\mu_1^2}{\mu^2}\right]$, for the two different
values of $\alpha$ pointed out above. We show also on each of
these
plots the value of the AMM calculated for the
$N=2$ truncation, which coincides with the AMM
obtained in the second order of perturbation theory. The results
for $\alpha = 0.2$ show rather good convergence as
$\mu_1\to \infty$. The contribution of the three-body Fock sector
to the AMM is sizeable but small, indicating
that the Fock decomposition~(\ref{Fock})
converges rapidly. This may show that once higher Fock components
are small, we can achieve practically converging calculation of
the AMM. Note that this value of $\alpha$ is not particularly
small: it is about 30 times the electromagnetic coupling, and is
about the size of the typical pion-nucleon coupling in  a nucleus.

When $\alpha$  increases, we see that the contribution of the
three-body sector considerably increases. For $\alpha = 0.5$  the
three-body contribution to the AMM starts to dominate at large
values of $\mu_1$. The dependence of the AMM on the PV boson mass
$\mu_1$ becomes more appreciable, 
although it keeps rather small.
\begin{figure}[ht!]
\begin{center}
\includegraphics[width=8.5cm]{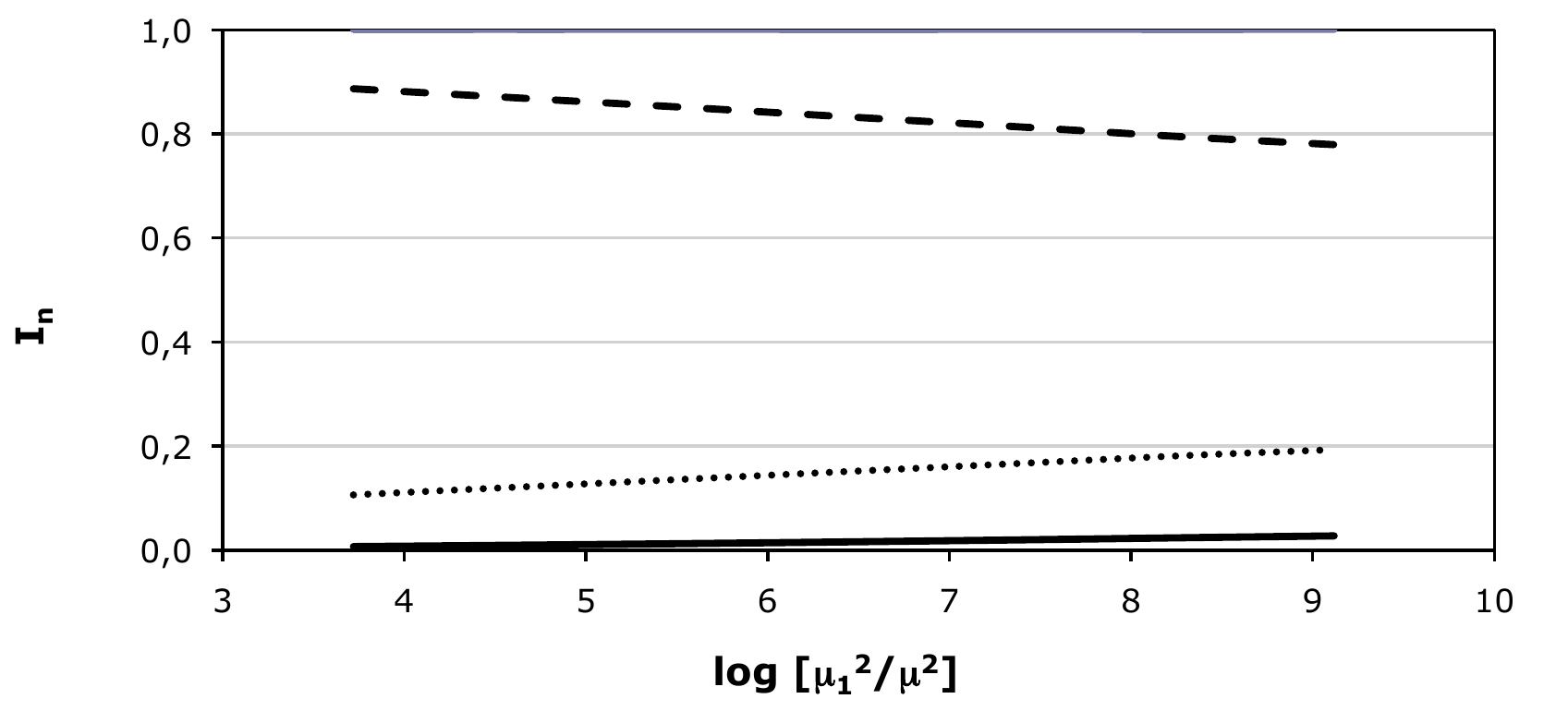}
\includegraphics[width=8.5cm]{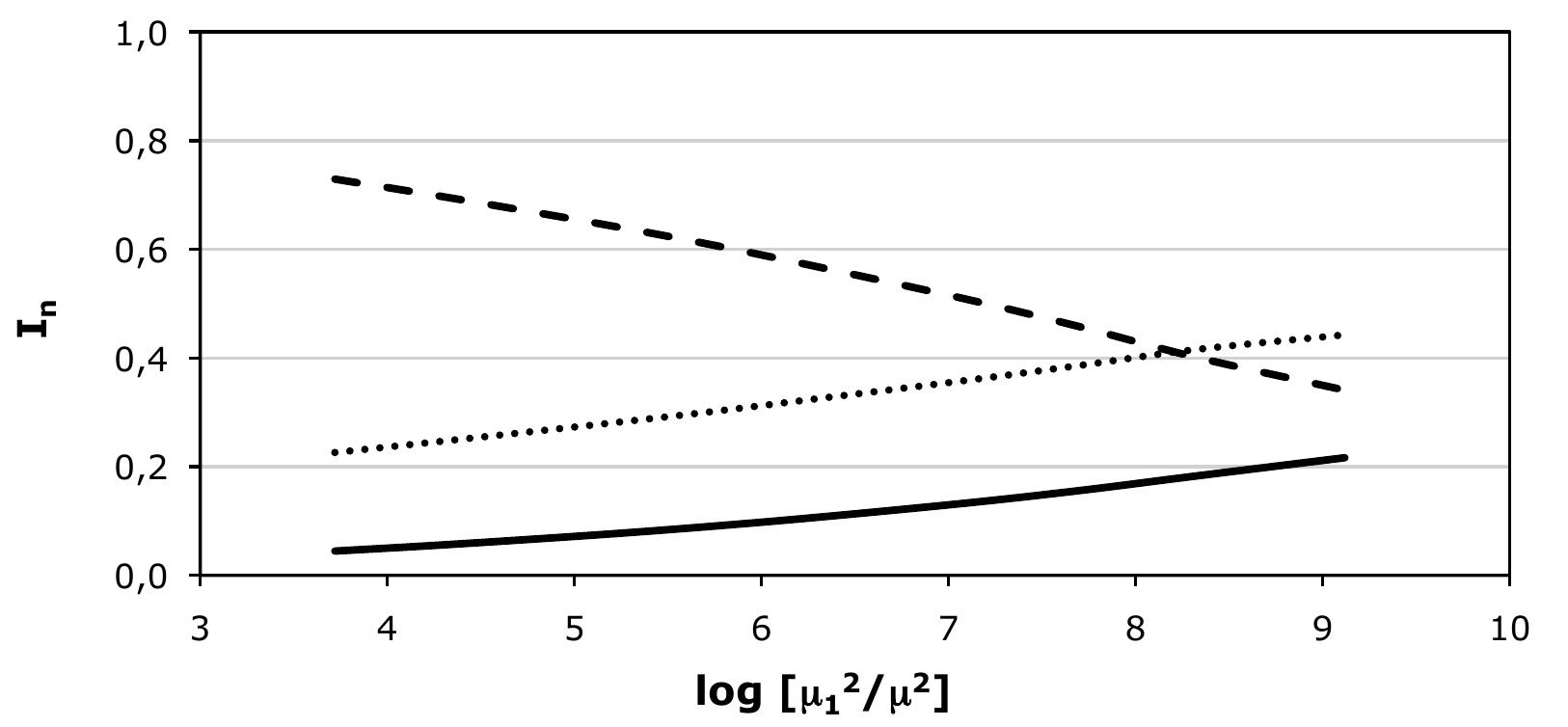}
\end{center}
\caption{Individual contributions of the one- (dashed line), two-
(dotted line), and three-body (solid line) Fock sectors to the
norm of the state vector as a function of the PV boson mass
$\mu_1$, for $\alpha =0.2$ (upper plot) and $\alpha = 0.5$ (lower plot).}\label{n123}
\end{figure}

In order to have a more physical insight into the relative
importance of different Fock sectors in the
decomposition~(\ref{Fock}) for the state vector, we plot in
Fig.~\ref{n123} the contributions of the one-, two-, and
three-body Fock sectors to the norm of the state vector for the
two values of the coupling constant, considered in this work. We
see again that at $\alpha=0.2$ the three-body
contribution to the norm is small, while it is not negligible, and
increases with $\mu_1$, when $\alpha = 0.5$.

At very large values of $\mu_1$, and for large $\alpha$, $I_1$ becomes negative. 
As already mentioned, we still get a well defined solution of Eqs.~(\ref{hH}), 
and there is no discontinuity whatsoever in the value of the AMM. As shown in Fig.~\ref{amm}, 
the convergence of the AMM as a function of the PV boson mass is expected in any case to 
settle much before we enter into this regime. According to renormalization theory, 
the mass of the PV boson should be much larger than any intrinsic momentum scale 
present in the calculation of physical observables. With this limitation, physical 
observables should be independent of any variation of the PV boson mass, within an 
accuracy which can be increased at will. This is what we found in our numerical 
calculation for small enough values of $\alpha$.
\begin{figure}[ht!]
\begin{center}
\includegraphics[width=8.5cm]{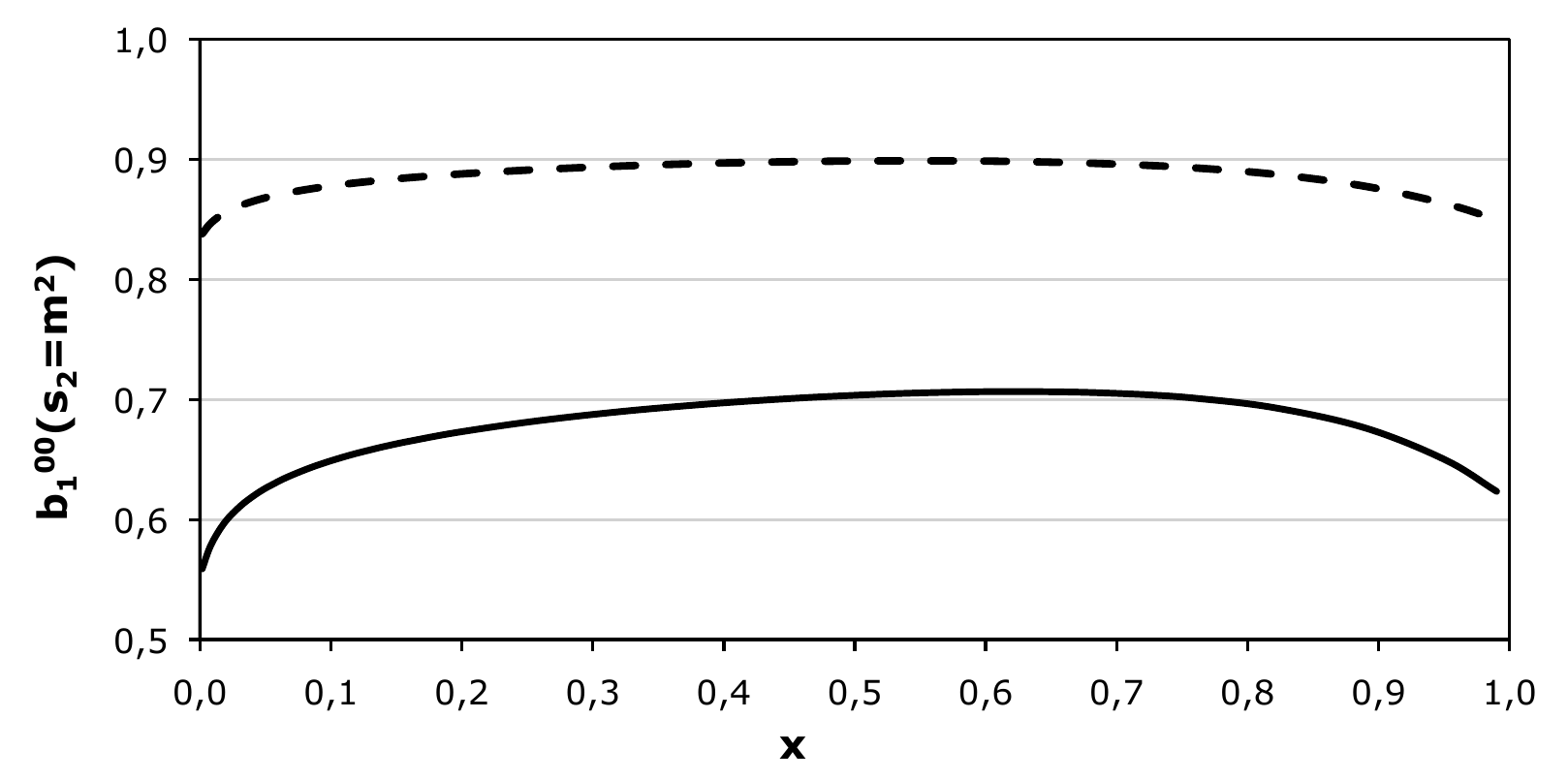}
\end{center}
\caption{The spin component $b_1^{00}$ of the two-body vertex
function~(\ref{gams}) calculated at $s_2=m^2$, as a function of
$x$, for $\alpha = 0.2$ (dashed line) and $\alpha = 0.5$ (solid
line), for a typical
value of $\mu_1=100$ GeV.}\label{b1on}
\end{figure}
\begin{figure}[ht!]
\begin{center}
\includegraphics[width=8.5cm]{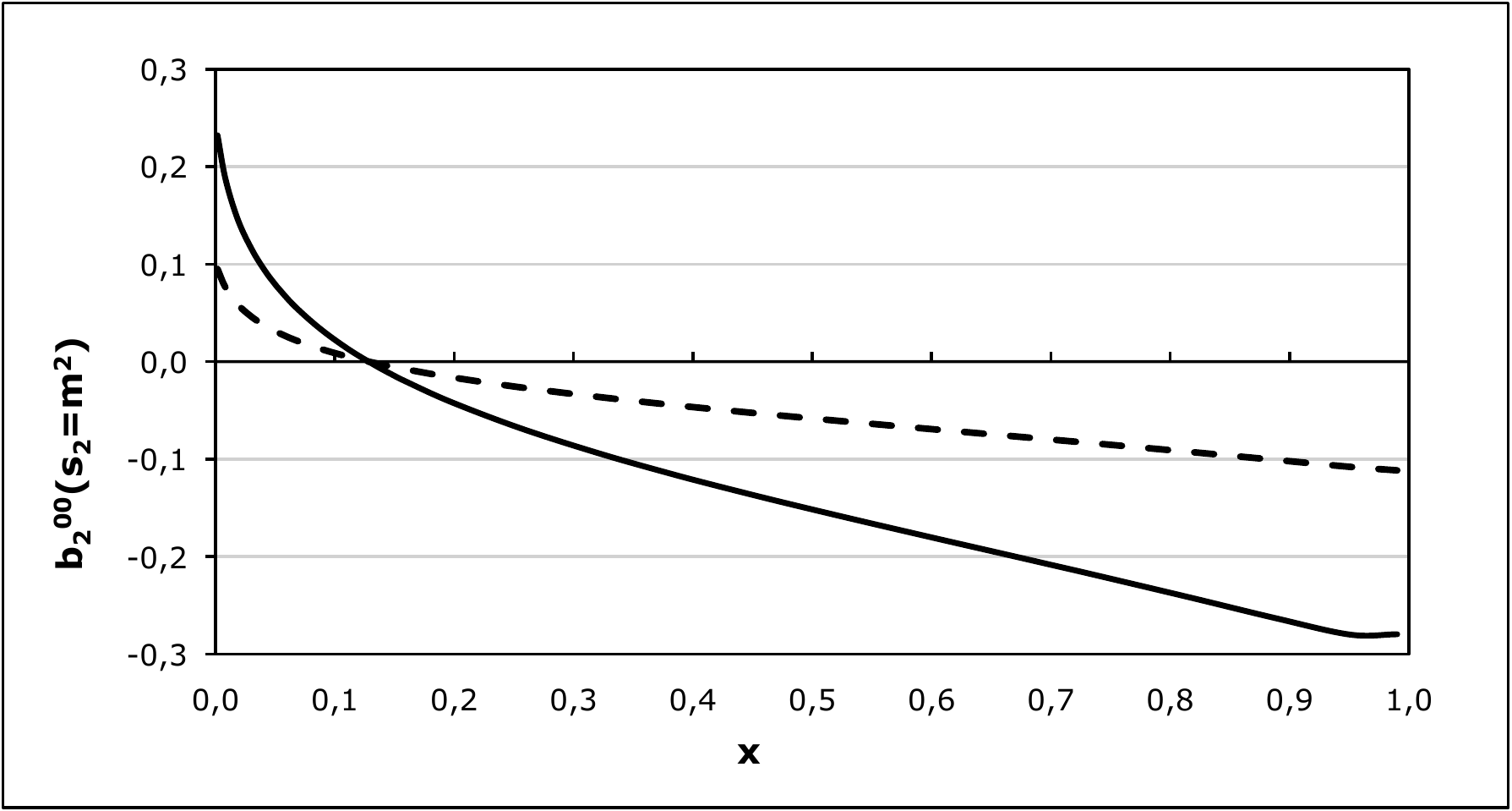}
\end{center}
\caption{The same as in Fig.~\ref{b1on},
 but for the  spin component $b_2^{00}$.}\label{b2on}
\end{figure}

In order to understand the possible origin of the residual
dependence of the AMM on $\mu_1$, we plot in Figs.~\ref{b1on}
and~\ref{b2on} the two-body spin components $b_1^{00}$ and
$b_2^{00}$ calculated at $s_2=m^2$, as a function of $x$. As we
already mentioned in Sec.~\ref{three-body}, $b_1^{00}(s_2=m^2)$
and $b_2^{00}(s_2=m^2)$ should be independent of $x$ in an exact
calculation. Moreover, $b_2^{00}$ should be zero. It is here fixed
to zero at a given value of $x=x^*\equiv\frac{\mu}{m+\mu}$, by the
adjustment of the constant $Z'_\omega$ in the system of
equations~(\ref{hH}). We clearly see in these figures that
$b_1^{00}$ is not a constant, although its dependence on $x$ is
always weak, while  $b_2^{00}$ is not identically zero, although
its value is relatively smaller than that of $b_1^{00}$ for
$\alpha=0.2$, and starts to be not negligible for $\alpha = 0.5$.

\begin{figure}[ht!]
\begin{center}
\includegraphics[width=8.5cm]{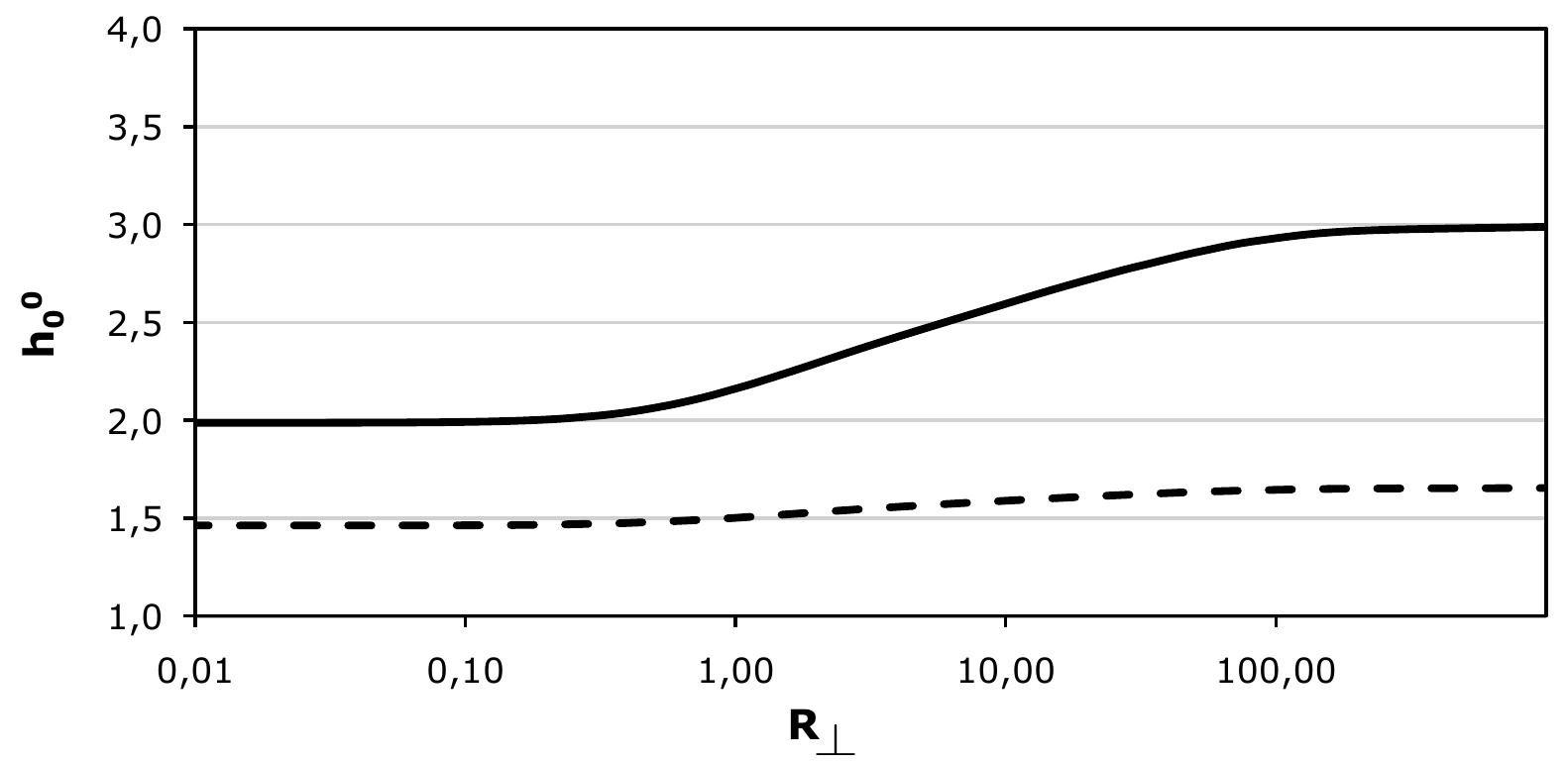}
\end{center}
\caption{The component $h^0_0$ defined by Eq.~(\ref{func}), as a
function of $R_\perp$ at $x=\frac{\mu}{\mu+m}$, for a typical
value of $\mu_1=100$ GeV and  for $\alpha = 0.2$ (dashed line) and
$\alpha = 0.5$ (solid line).}\label{h0}
\end{figure}
\begin{figure}[ht!]
\begin{center}
\includegraphics[width=8.5cm]{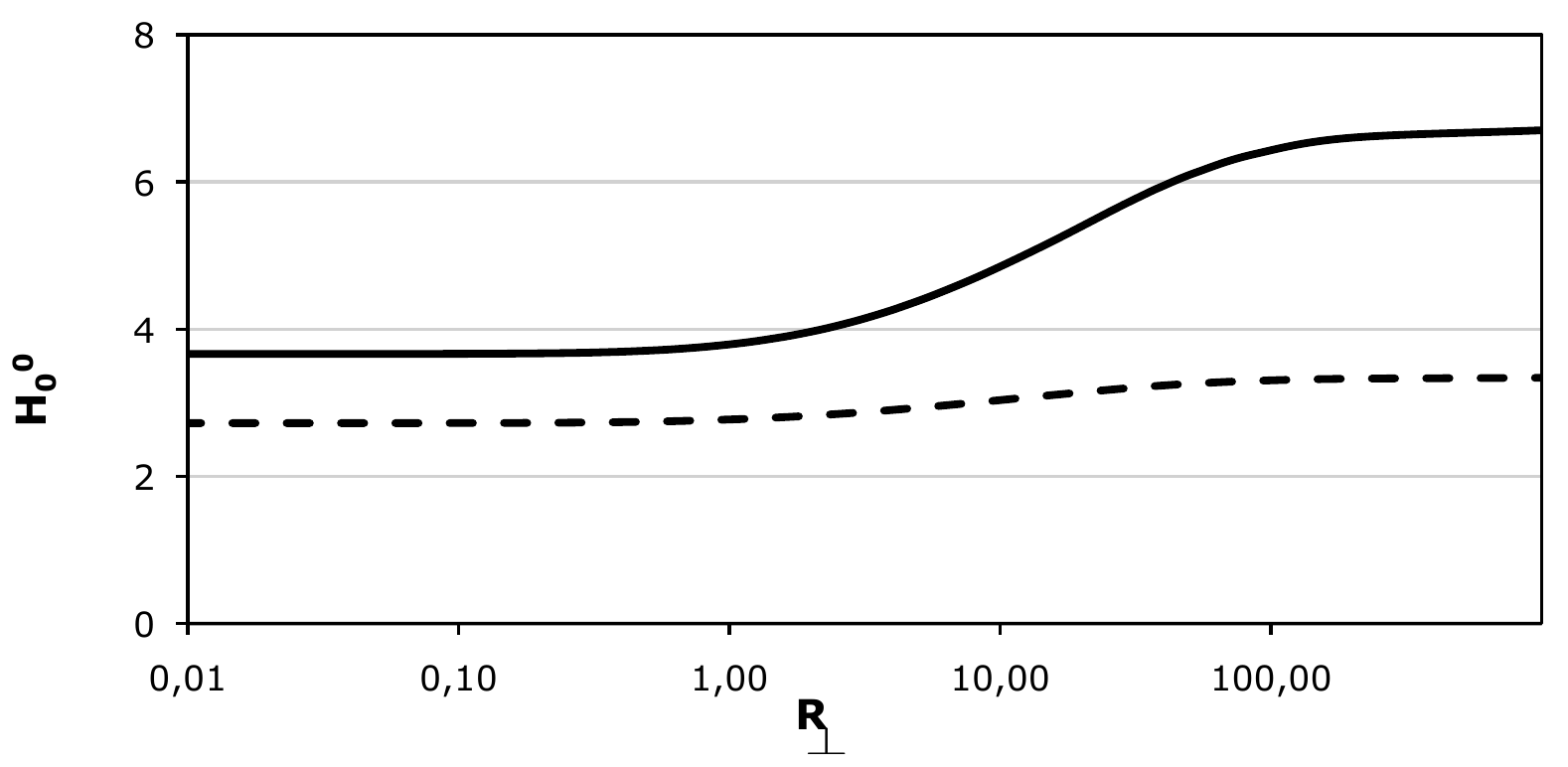}
\end{center}
\caption{The same as in Fig.~\ref{h0}, but for the component
$H_0^0$.\label{HH0}}
\end{figure}

We plot in  Figs.~\ref{h0} and~\ref{HH0} the two physical
components $h_0^0$ and $H_0^0$ as a function of $R_\perp$, at
$x=\frac{\mu}{m+\mu}$. As expected from the system of eigenvalue
equations~(\ref{hH}),
the functions $\tilde{h}_0^0$
and $\tilde{H}_0^0$ tend to constants at large $R_\perp$.
Hence, the functions ${h}_0^0$ and ${H}_0^0$
related to them by Eq.~(\ref{func}) tend to constants too. Note
that the two-body wave function $\phi_2$ related to $\Gamma_2$ by
Eq.~(\ref{Gn}) goes to zero at large momenta due to the rapidly
decreasing kinematical factor $(s_2-m^2)^{-1}$.

\begin{figure}[ht!]
\begin{center}
\includegraphics[width=8.5cm]{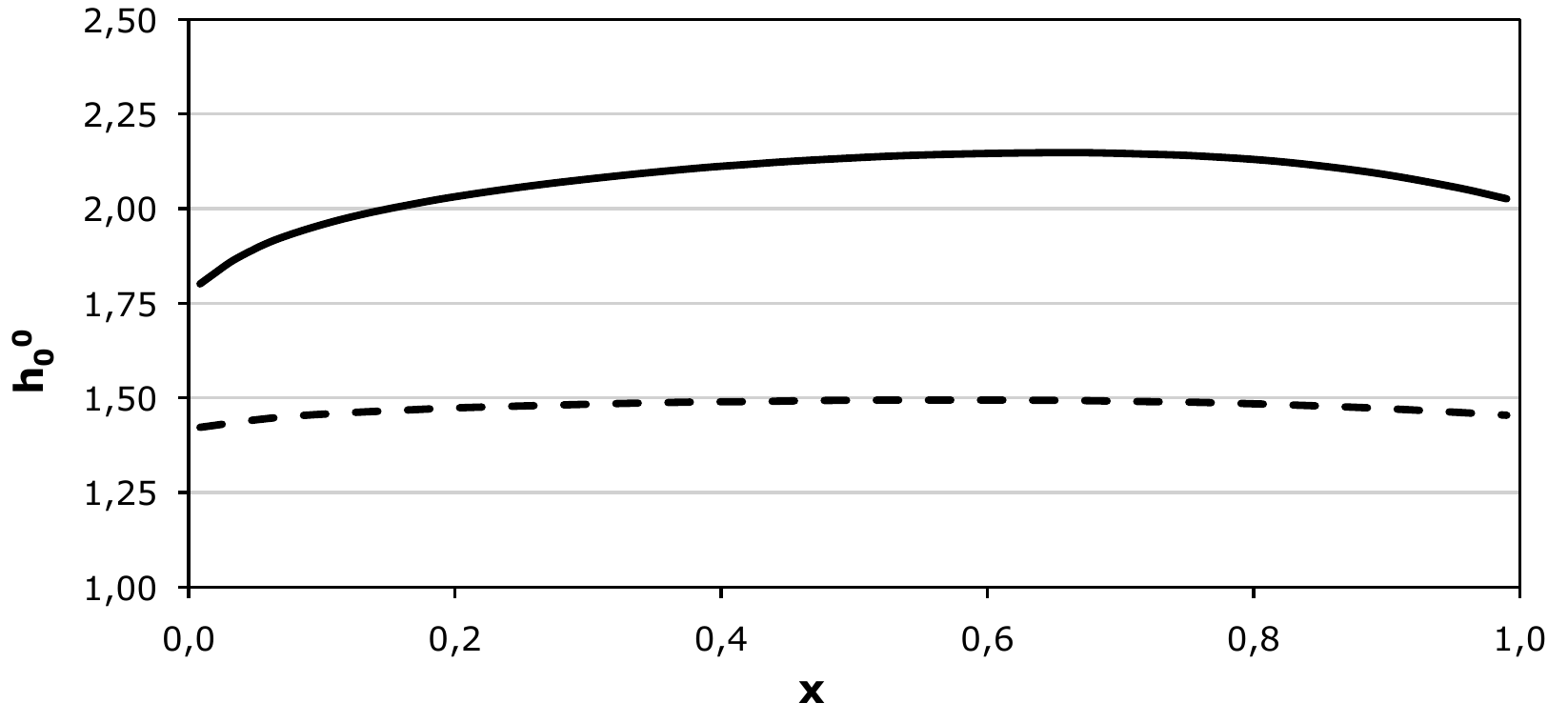}
\end{center}
\caption{The component $h^0_0$ defined by Eq.~(\ref{func}), as a
function of $x$ at $R_\perp = 0$, for a typical
value of $\mu_1=100$ GeV and  for $\alpha = 0.2$ (dashed line) and
$\alpha = 0.5$ (solid line).}\label{h0x}
\end{figure}

\begin{figure}[ht!]
\begin{center}
\includegraphics[width=8.5cm]{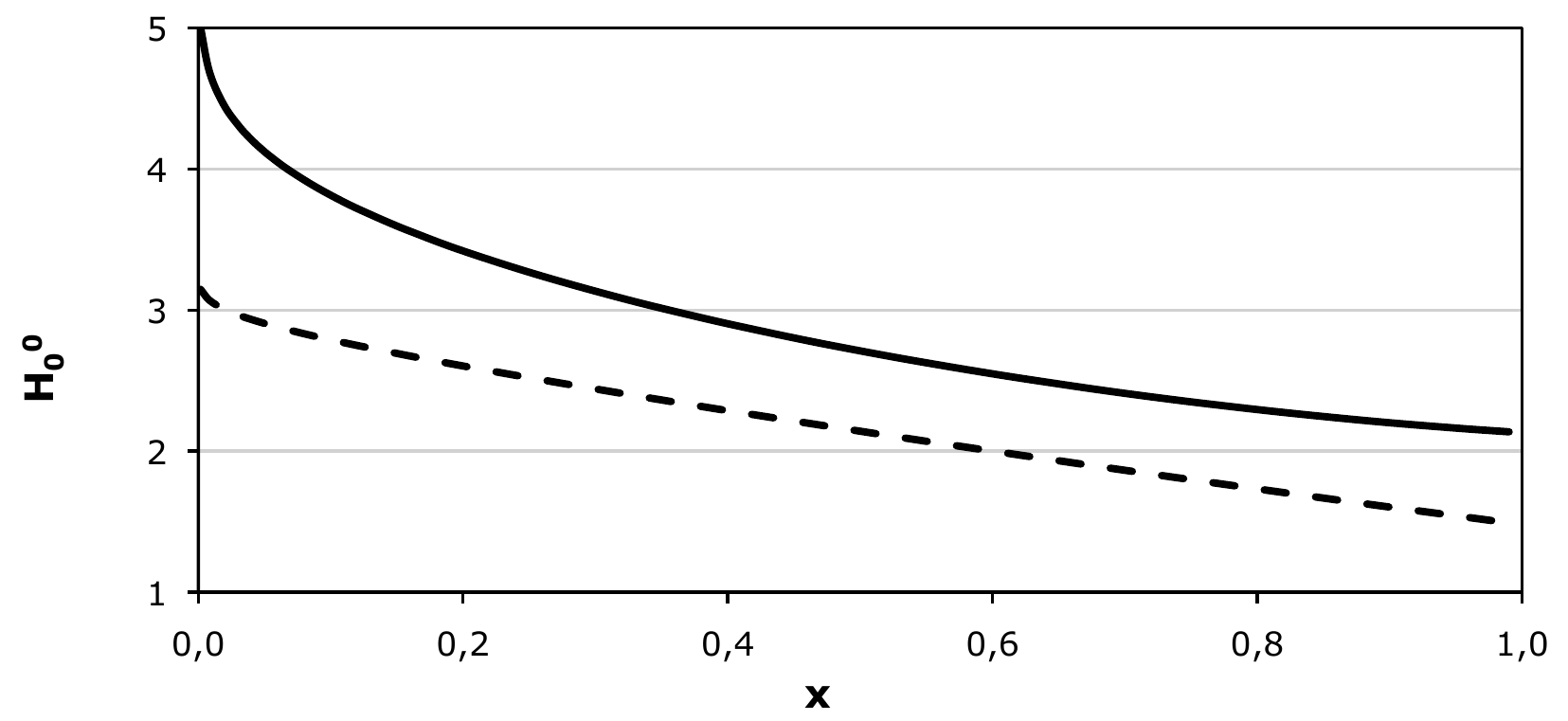}
\end{center}
\caption{The same as in Fig.~\ref{h0x}, but for the component
$H_0^0$.}\label{HH0x}
\end{figure}

For completness, we plot in
Figs.~\ref{h0x} and~\ref{HH0x} the two physical
components $h_0^0$ and $H_0^0$ as a function of $x$, at $R_\perp =
0$.

%

\section{Conclusion} \label{conc}
We calculated the anomalous magnetic moment of a fermion in the
Yukawa model, in the first non-trivial approximation,
incorporating a constituent fermion coupled to zero, one, and two
scalar bosons, i.e.
within the three-body Fock state truncation. We
applied a general formalism based on the covariant formulation of
light-front dynamics and an appropriate Fock sector dependent
renormalization scheme which enables us to control uncancelled
divergences when Fock space is truncated. We paid particular
attention to the renormalization conditions necessary to relate
the bare coupling constant to the physical one. To do this, we
need to identify all spurious contributions originating from the
violation of rotational invariance, coming from the Fock space
truncation. This is possible in our covariant formulation.

The anomalous magnetic moment shows a very nice convergence as a
function of the regularization scale (the Pauli-Villars boson mass
$\mu_1$), for the coupling constant value $\alpha = 0.2$ which
mimics a nucleon coupled to a scalar "pion". For this value of
$\alpha$, the two-body component gives a dominant contribution to
the anomalous magnetic moment. As $\alpha$ increases, we see the
onset of higher Fock components.

This shows up in the large contribution of the three-body
component, and in the dependence of the anomalous magnetic moment
as a function of the regularization scale. We believe that this
latter dependence should be largely, if not completely, removed by
incorporating the relevant fermion-antifermion contributions to
the three-body Fock components. We are presently investigating
these contributions~\cite{kms09}.

As we have seen in our study, the calculation of nonperturbative
properties of bound state systems demand to control all
approximations in a quantitative way, in order to be able to
make physical predictions order by order in the Fock space truncation. 
We think that the combination of the covariant  formulation of light-front dynamics
with an appropriate Fock sector dependent renormalization scheme
is  a quite promising method to investigate these properties in a
very elegant way, with a minimum of Fock components and
computational time.

\begin{acknowledgments}
Two of us (V.~A.~K. and A.~V.~S.) are sincerely grateful for the
warm hospitality of the Laboratoire de Physique Corpusculaire,
Universit\'e Blaise Pascal, in Clermont-Ferrand, where part of
the present study was performed.
\end{acknowledgments}

\appendix
\section{Relation between the Fock component normalization
and the field strength renormalization factors} \label{renorma}

We shall prove here the relation~(\ref{IZ}) in the general case,
i.~e. without Fock space truncation. We omit for a moment
antifermion contributions generated by the process $b\to
f\bar{f}$. They will be incorporated below. The fermion
self-energy is given by a sum of irreducible graphs with all
possible intermediate states:
\begin{equation}
\label{Sigs} \Sigma({\sla p})=\sum_{n=1}^{\infty} \Sigma_n({\sla
p}).
\end{equation}
Here $\Sigma_n({\sla p})$ is the contribution from a graph  with
$n$ intermediate states. For example, the first
graph in Fig.~\ref{fig4} for the self-energy
contains
one intermediate state, the second one contains
three intermediate states while the third one
contains 11 intermediate states.

For the calculations we will use the 3D light-front graph
techniques exposed in Ref.~\cite{cdkm}. It can be represented in
different equivalent forms. We use the form in which an
amplitude is represented as a product of energetic denominators
(one denominator for each intermediate state) multiplied by
appropriate spin matrices. The amplitude (with a conventional minus
sign for the self-energy) corresponding to a graph
with $n$ intermediate states has the following form:
\begin{equation}
\label{Sign}
 \Sigma_n({\sla p})=-g_0^{n+1}\int dD\;\;
\prod_{i=1}^n\frac{{\sla k}_i+m_i}{s_i-p^2}.
\end{equation}
Since we do not truncate Fock space, we use the same BCC in all
vertices. $s_i$ is the square of the invariant energy of $i$-th
intermediate state. The product $\prod_{i=1}^n$ runs through all
$n$ intermediate states. The integration in Eq.~(\ref{Sign}) is
performed over all independent variables.

The decomposition of $\Sigma({\sla p})$ is similar to that given
by Eq.~(\ref{Sigk}):
\begin{equation}
\label{Sigp} \Sigma({\sla p}) = {\cal A}(p^2)+ {\cal B}(p^2)
\frac{\sla p}{m} + {\cal C}(p^2)\; \frac{m{\sla
\omega}}{\omega\cd p}+{\cal C}_1(p^2)\sigma,
\end{equation}
where
$$
\sigma=\frac{1}{4\omega\cd p}({\sla p}{\sla \omega}-{\sla
\omega}{\sla p}).
$$
Eq.~(\ref{Sigp}) is the most general form of the fermion
self-energy in CLFD. The term with the function $C_1(p^2)$ does
not appear for the two-body ($N=2$) Fock space truncation, but
it may appear for $N\geq 3$. We give here the coefficient
${\cal C}$ which will be used below:
\begin{equation}
\label{C} {\cal C} =\frac{1}{4m}\mbox{Tr}\left[\left({\sla
p}-{\sla \omega}\frac{p^2}{\omega\cd p}\right)\Sigma({\sla
p})\right].
\end{equation}

To find  $\left.\frac{\partial \Sigma({p\! \! \!/})}{\partial {p\! \! \!/}}\right\vert_{p \! \! \!/ = M}$, we first replace in
Eq.~(\ref{Sigp}) $p^2$ by $M^2$, ${\sla p}$ by $M$ (that is, replace
$\gamma^{\nu}$ by $p^{\nu}/M$) and then calculate the derivative
over $M$. It is convenient to make this replacement by using the
formula
$$
\Sigma(M) = \frac{1}{4M}\mbox{Tr}\left[({\sla p}+M)\Sigma({\sla
p})\right]_{p^2=M^2}.
$$
We get
\begin{equation}
\label{der} \left.\frac{\partial \Sigma({\sla p})}{\partial \sla
p}\right\vert_{p \! \! \!/ = M}=\frac{\partial}{\partial
M}\left\{\frac{1}{4M} \mbox{Tr}[({\sla p}+M)\Sigma({\sla
p})]_{p^2=M^2}\right\}.
\end{equation}
We substitute here $\Sigma({\sla p})$ from Eqs.~(\ref{Sigs})
and~(\ref{Sign}).

The contribution of derivative from the $j$-th factor
$\frac{1}{s_j-M^2}$ of the denominator, which results from
Eq.~(\ref{Sign}), reads
\begin{widetext}
\begin{equation}\label{Sigd2}
\left.\frac{\partial \Sigma^{den}_{nj}({\sla p})}{\partial \sla
p}\right\vert_{{p\! \! \!/} = M}= -g_0^{n+1}2M\int
dD\;\mbox{Tr}\left\{\frac{1}{4M}({\sla p}+M)
\left[\prod_{i_1=1}^{j-1}\frac{({\sla
k}_{i_1}+m_{i_1})}{(s_{i_1}-M^2)}\right]\; \frac{({\sla
k}_{j}+m_{j})}{(s_j-M^2)^2}\;
\left[\prod_{i_2=j+1}^{n}\frac{({\sla
k}_{i_2}+m_{i_2})}{(s_{i_2}-M^2)}\right]\right\}.
\end{equation}
\end{widetext}
The factor
\begin{equation}\label{Gamma}
 \Gamma_{j} = g_0^j\int dD' \;
\prod_{i_1=1}^{j-1}\frac{({\sla
k}_{i_1}+m_{i_1})}{(s_{i_1}-M^2)}
\end{equation}
is a contribution of the graph with $j-1$ intermediate states into
the vertex function, and similarly for the second product. In
contrast to Eq.~(\ref{Sigd2}), where the integration $dD$ runs
over the phase volumes of all the intermediate states, the integration $dD'$ in
Eq.~(\ref{Gamma}) runs over the phase
volumes of the intermediate states $i_1=1,\ldots,j-1$ only.

Since all the four-momenta are on the
corresponding mass shells $k_j^2=m_j^2$, we have
\begin{eqnarray*}
({\sla k}_{j}+m_{j})&=&\sum_{\sigma=\pm
1/2}u_{\sigma}(k_j)\bar{u}_{\sigma}(k_j),
\\
\frac{1}{2}\mbox{Tr}[({\sla
p}+M)O]&=&\frac{1}{2}\sum_{\lambda=\pm 1/2}\bar{u}_{\lambda}(p)O
u_{\lambda}(p),
\end{eqnarray*}
for an arbitrary matrix $O$. The factor
$\frac{1}{2}$ in the last equation is introduced for averaging
over initial spin projections.

We therefore get
\begin{widetext}
\begin{equation}\label{dSig}
\left.\frac{\partial \Sigma^{den}_{nj}({\sla p})}{\partial \sla
p}\right\vert_{{p\! \! \!/} = M}=
-\frac{1}{2}\sum_{\lambda,\sigma}\int dD_j
\frac{\bar{u}_{\lambda}(p)\Gamma_{j}u_{\sigma}(k_j)}{(s_j-M^2)}
\frac{[\bar{u}_{\lambda}(p)\Gamma_{j}u_{\sigma}(k_j)]^{\dagger}}{(s_j-M^2)}=
-\frac{1}{2}\sum_{\lambda,\sigma}\int dD_j\;
\phi^{\lambda}_{j,\sigma}(p)\phi^{\lambda\dagger}_{j,\sigma}(p)
\end{equation}
\end{widetext}
Here the integration $dD_j$ runs over the phase volume of the $j$-th
intermediate state not included in the integral for $\Gamma_j$.
The vertex function $\Gamma_j$ may correspond to any fixed
number of particles in the state $j$ allowed by a given graph.
We took into account that the factor $1/(s_j-M^2)$ turns each
$\Gamma$ into $\phi$, according to Eq.~(\ref{Gn}). Taking the
sum over all the graphs and over all the intermediate states
$j$, we recover in Eq.~(\ref{dSig}) 
the contribution to the normalization integral $I_{N\geq 2}$
from all the $N$-body states with $N\geq 2$
(each intermediate state
in irreducibles graphs for $\Sigma$ contains at least two particles).
Since the rules of the graph
techniques used to calculate  $\Sigma$ imply that the one-body
states are normalized to $1$, this means that $I_{N\geq 2}$
corresponds to a state vector normalized by the condition $I_1=1$.
If $I_1\neq 1$, then Eq.~(\ref{Sigd2}) determines the ratio
$-I_{N\geq 2}/I_1$. That is
\begin{equation}
\label{denr} \left.\frac{\partial \Sigma^{den}({\sla
p})}{\partial \sla p}\right\vert_{p \! \! \!/ =
M}=-\frac{I_{N\geq 2}}{I_1}.
\end{equation}

The contribution of the derivative of other factors in
Eq.~(\ref{der}), except for $\prod_i\frac{1}{s_i-M^2}$, reads
\begin{widetext}
\begin{equation}
\label{Sign1} \left.\frac{\partial \Sigma^{num}({\sla
p})}{\partial \sla p}\right\vert_{{p\! \! \!/}= M}=-g_0^{n+1}\int
dD\; \frac{\frac{\partial}{\partial M}\left\{\frac{1}{4M}
\mbox{Tr}[({\sla p}+M)\prod_{i=1}^n({\sla
k}_i+m_i)]_{p^2=M^2}\right\}} {\prod_{i=1}^n(s_i-M^2)}.
\end{equation}
\end{widetext}
Consider first the case when the products in Eq.~(\ref{Sign1})
contain only one factor with a fixed $i$. Then $\Sigma$
corresponds to the first graph in Fig.~\ref{fig4}. We calculate
the trace in Eq.~(\ref{Sign1}), using the following explicit
expression for the scalar product:
$$
k_i\cd p=\frac{1}{2x_i}({\bf R}_{i{\perp}}^2 + m_i^2+ x_i^2M^2),
$$
where the variables ${\bf R}_{i{\perp}}$ and $x_i$ are constructed
according to Eq.~(\ref{lf}). Calculating then the derivative over
$M$, we find
\begin{widetext}
\begin{equation}
\label{num1} \left.\frac{\partial \Sigma^{num}({\sla
p})}{\partial \sla p}\right\vert_{{p\! \! \!/}= M}=-g_0^2\int dD_2
\frac{\frac{\partial}{\partial M}\left\{\frac{1}{4M}
\mbox{Tr}[({\sla p}+M)({\sla k}_i+m_i)]\right\}}{(s_i-M^2)}=
-g_0^2 \int d D_2 \frac{\frac{1}{2x_i M^2}({\bf R}_{i{\perp}}^2
+ m_i^2- x_i^2M^2)}{(s_i-M^2)},
\end{equation}
\end{widetext}
where $dD_2=\frac{d^2R_{{i}\perp}dx_{i}}
{(2\pi)^3\,2x_{i}(1-x_{i})}$ is the two-body phase space volume
element.

Let us calculate now the value of the coefficient ${\cal C}$ in
Eq.~(\ref{Sigp}). It is given by Eq.~(\ref{C}). We still
consider a particular case and keep one factor only with fixed
$i$. Then ${\cal C}$ obtains the form
\begin{widetext}
\begin{equation}
\label{C1} {\cal C}=-g_0^2\int dD_2\frac{
\frac{1}{4m}\mbox{Tr}\left[({\sla
p}-\sla{\omega}\frac{p^2}{\omega\cd p})({\sla
k}_i+m_i)\right]}{(s_i-M^2)}= -g_0^2\int dD_2
\frac{\left(-\frac{M^2}{m}\right)\frac{1}{2x_i M^2}({\bf
R}_{i{\perp}}^2 + m_i^2- x_i^2M^2)}{(s_i-M^2)}.
\end{equation}
\end{widetext}
Comparing Eq.~(\ref{num1}) with Eq.~(\ref{C1}), we find the
relation
\begin{equation}
\label{num2} \left.\frac{\partial \Sigma^{num}({\sla
p})}{\partial \sla p}\right\vert_{{p\! \! \!/}=
M}=-\frac{m}{M^2}{\cal C}.
\end{equation}

It turns out that Eq.~(\ref{num2}) is valid in the most general case. In
the latter case, but still without antifermions, we get in the
numerator in Eqs.~(\ref{num1}) and (\ref{C1}) a product of the
matrices $\prod_{i=1}^n({\sla k}_i+m_i)$, instead of the single
term $({\sla k}_i+m_i)$. This product can be decomposed in the
full set of the $4\times 4$ matrices as follows:
\begin{widetext}
\begin{equation}
\label{decomp} \prod_{i=1}^n({\sla k}_i+m_i)=G_0+\sum_i G^i_1
{\sla k}_i +\sum_{i_1, i_2} G^{i_1 i_2}_2\;
\sigma(k_{i_1},k_{i_2}) + G_3 \gamma_5+ \sum_iG^i_4
\gamma_5{\sla k}_i,
\end{equation}
\end{widetext}
where $\sigma(k_{i_1},k_{i_2})=\frac{i}{2}({\sla k}_{i_1}{\sla
k}_{i_2}- {\sla k}_{i_2}{\sla k}_{i_1})$. The coefficients
$G_{1-3}$ depend on the scalar products of the four-momenta
$k_1,\ldots k_n$:
$$
k_{i}\cd k_j=\frac{1}{2x_i x_j}[x_i^2 m_j^2 +(x_j{\bf
R}_{i\perp}- x_i{\bf R}_{j\perp})^2 +x_j^2 m_i^2].
$$
It is important that these scalar products and, hence, the
functions $G_{1-3}$ do not depend on $M$. Therefore $G_{1-3}$
can be extracted from the operator $\frac{\partial}{\partial
M}$. We replace $({\sla k}_i+m_i)$ in Eqs.~(\ref{num1})
and~(\ref{C1}) by the product $\prod_{i=1}^n({\sla k}_i+m_i)$
represented in the form~(\ref{decomp}) (and take the product of
the denominators). The matrices $\gamma_5$ and $\gamma_5{\sla
k}_i$ give zero contributions to both Eqs.~(\ref{num1})
and~(\ref{C1}), whereas with the matrices 1, ${\sla k}_i$, and
$\sigma(k_{i_1},k_{i_2})$ we reproduce the
relation~(\ref{num2}).

The incorporation of antifermions (say, the $ff\bar{f}$
intermediate state, in addition to bosons)
does not change the form of the denominator (though the energies
$s_i$ incorporate now the antifermion momenta). That results in
Eq.~(\ref{denr}). The corresponding numerator contains now the
spin matrices of all the fermions + antifermions [we get one
factor $1/(s_i-p^2)$ and a product of three matrices $(\pm {\sla
k}_i+m_i)$ for the $ff\bar{f}$ state; the signs plus and
minus stand for fermions and antifermions, respectively]. We
still can decompose the full numerator according to Eq.~(\ref{decomp}) and
again reproduce the formula~(\ref{num2}).

In this way, taking the sum of Eqs.~(\ref{denr})
and~(\ref{num2}), we finally find
\begin{equation}
\label{dsig} \left.\frac{\partial \Sigma({\sla p})}{\partial
\sla p}\right\vert_{{\sla p}= M}  = -\frac{I_{N\geq 2}}{I_1}
-\frac{m}{M^2}{\cal C}.
\end{equation}
If rotational invariance is preserved (it can be violated for instance by omitting some
time-ordered graphs or by using rotationally non-invariant cutoffs), ${\cal C}$
is zero. It is indeed zero, for example, in the two-body
approximation with the PV regularization [see Eq.~(\ref{CC}) in
Appendix~\ref{secoef}]. If ${\cal C}=0$, substituting
Eq.~(\ref{dsig}) into Eq.~(\ref{Zf}) and taking into account that
$I_1+I_{N\geq 2}=1$, we prove the relation~(\ref{IZ}).

\section{Self-energy coefficients} \label{secoef}
We give here explicit formulas for the coefficients ${\cal
A}(k^2)$, ${\cal B}(k^2)$, and ${\cal C}(k^2)$ entering
Eq.~(\ref{Sigk}) for the two-body self-energy in the Yukawa model.
If $\Sigma({\sla k})$ is known, these coefficients can be found as
follows:
\begin{eqnarray}
\label{ABC1} g_{02}^2\ {\cal A}(k^2)&=&\frac{1}{4}
\mbox{Tr}\left[\Sigma({\sla k})\right],\\
\label{ABC2} g_{02}^2\ �{\cal B}(k^2)&=&\frac{m}{4 \omega\cd k}
\mbox{Tr}\left[\Sigma({\sla k}){\sla
\omega}\right],\\
\label{ABC3} g_{02}^2\  {\cal
C}(k^2)&=&\frac{1}{4m}Tr\left[\Sigma({\sla k})\left({\sla
k}-\sla{\omega}\frac{k^2}{\omega\cd k}\right)\right].
\end{eqnarray}

In the Yukawa model, the self-energy regularized by one PV boson
and one PV fermion reads, in CLFD,
\begin{multline}
\label{se1} \Sigma({\sla k})=-\frac{g_{02}^2}{(2\pi)^3}\int
d^2R_{\perp}\int_0^1
\frac{dx}{2x(1-x)} \\
\times\sum_{i,j=0}^1 (-1)^{i+j} \frac{({\sla
k}_1+m_i)}{(s_{ij}-k^2)},
\end{multline}
where $k_1$ is the internal fermion four-momentum. The
light-front variables are, as usual, ${\bf R}_{\perp}={\bf
k}_{2\perp}-x{\bf k}_{\perp}$,  $x=\omega\cd k_2/\omega\cd k$
($k_2$ is the
boson four-momentum), and
\begin{equation} \label{s}
s_{ij}=\frac{{R}_{\perp}^2+\mu_j^2}{x}+\frac{{
R}_{\perp}^2+m_i^2}{1-x}.
\end{equation}

The coefficients
${\cal A}$ and ${\cal B}$ converge without PV fermion (i.~e., they
have finite limit when $m_1\to\infty$).
Substituting Eq.~(\ref{se1}) into
Eqs.~(\ref{ABC1}) and~(\ref{ABC2}), integrating over $d^2R_{\perp}$ and
omitting the PV fermion, we get
\begin{widetext}
\begin{eqnarray*}
{\cal A}({k}^2)= \frac{m}{16\pi^2}\int_0^1 \log\left[
\frac{xm^2-x(1-x){k}^2+(1-x)\mu^2}{xm^2-x(1-x){k}^2+(1-x)\mu_1^2}\right]dx,&&
\\
{\cal B}({k}^2)=\frac{m}{16\pi^2}\int_0^1 (1-x) \log\left[
\frac{xm^2-x(1-x){k}^2+(1-x)\mu^2}{xm^2-x(1-x){k}^2+(1-x)\mu_1^2}\right]dx.&&
\end{eqnarray*}
\end{widetext}
Notice that in the limit $\mu_1\to\infty$ and
for fixed
${k}^2$ the values of  ${\cal A}_r({k}^2)=
{\cal A}({k}^2) -{\cal
A}(m^2)$ and ${\cal B}_r({k}^2)={\cal B}({k}^2) -{\cal B}(m^2)$ are
finite.

A similar calculation of ${\cal C}({k}^2)$ requires, for
convergence, not only one PV boson, but also one PV fermion. We thus
find
\begin{widetext}
\begin{equation}\label{CC}
{\cal C}({k}^2)=-\frac{1}{32 m \pi^2}\int_0^1\frac{dx}{1-x}
\int_0^{\infty}dR^2_{\perp} 
\sum_{i,j=0}^1(-1)^{i+j}  \frac{{ R}^2_{\perp}-(1-x)^2{k}^2+m_i^2}
{R^2_{\perp}-x(1-x){k}^2+(1-x)\mu_j^2+xm_i^2}\equiv 0.
\end{equation}
\end{widetext}
%

\section{Right-hand sides of the eigenvalue equations~(\ref{eq245})} \label{V16}
The system of equations~(\ref{eq245}) determines the one- and
two-body Fock components $\Gamma_{1}^{i}$, $\Gamma_{2}^{ij}$. The
r.-h.~s.'s of these equations are denoted by
$\bar{u}(p_{1i})\left(V_{1}+V_{2}\right)u(p)$ and
$\bar{u}(k_{1i})\left(V_{3}+V_{45}+V_{6}\right)u(p)$,
respectively. The explicit form of $V_{1-6}$ is the following:

\begin{subequations}
\begin{widetext}
\begin{eqnarray}
\label{V12b} V_{1} & = & {\delta
m_3}\sum_{i'}(-1)^{i'}\frac{({\sla p_{i'}}+m_{i'})}{m_{i'}^2-M^2} \Gamma_{1}^{i'},
\\
\label{V22b}
 V_{2}& = &
{g_{03}'}\sum_{i',j'}(-1)^{i'+j'}\int
\frac{d^2R'_{\perp}}{(2\pi)^3}
 \int_{0}^{1}\frac{dx'}{2x'}\,
 \frac{({\sla k}'_{1i'}+m_{i'})}{2(\omega\cdot p)\tau_{i'j'}}
\Gamma_{2}^{i'j'},
 \nonumber\\
\label{V32b} V_{3} & = & {g_{03}'}\sum_{i'}(-1)^{i'}
\frac{({\sla p_{i'}}+m_{i'})}{m_{i'}^2-M^2} \Gamma_{1}^{i'}, \\
\label{V45} V_{45} & = & {\left[-\Sigma({\sla p}-{\sla
k}_{2j})+\delta m_2\right]} \sum_{i'}(-1)^{i'} \frac{({\sla
k}'_{1i'}+m_{i'})}{2(\omega\cdot p)(1-x)\tau_{i'j}}
\Gamma_{2}^{i'j},\\
 V_{6} & = &
 g_{02}^2\sum_{i',j',i''}(-1)^{i'+j'+i''}\int
\frac{d^2R'_{\perp}}{(2\pi)^3}\int_{0}^{1-x} \frac{dx'}{2x'}\nonumber \\
&&{\times} \frac{\left({\sla k}''_{1i''}+m_{i''}\right)
\left({\sla k}'_{1i'}+m_{i'}\right)} {4(\omega\cdot
p)^2(1-x')(1-x-x')\tau_{i'j'}\tau_{i''jj'}} \Gamma_{2}^{i'j'}
\nonumber \\
\label{V6}
\end{eqnarray}
\end{widetext}
\end{subequations}
with $g_{03}'=g_{03}+Z_{\omega}\frac{m{\sla \omega}}{\omega\cd
p}$ and obvious notations for the momenta of the particles in the intermediate states.
The term $V_{45}$ stands for the sum of the contributions of
the graphs $V_4$ and $V_5$ in Fig.~\ref{fig4}. The two-body vertex functions $\Gamma_2$ inside the integrands
depend on ${\bf R}'_{\perp}$ and $x'$, while those which are not
integrated depend on ${\bf R}_{\perp}$ and $x$. After calculating
the traces taken from the equations~(\ref{eq24p})
and~(\ref{eq25p}), we obtain scalar products which are
expressed through the variables ${\bf R}_{\perp},x$ and ${\bf
R}'_{\perp},x'$. Examples are given in appendix C of
Ref.~\cite{kms_08}.

The values of $\tau$'s, which appear in the above formulas, are
related to the invariant energies in the corresponding
intermediate states. For example, $\tau_{i''jj'}$ in Eq.~(\ref{V6})
for $V_6$ has the form
$$
\tau_{i''jj'}=\frac{s_{i''jj'}-m^2}{2\omega\cd p} ,
$$
where
$$
s_{i''jj'}=(k_{1i''}+k_{2j}+k'_{2j'})^2.
$$
$k_{1i''}$, $k_{2j}$, and $k'_{2j'}$ are the four-momenta in the
intermediate states while
$s$, for any intermediate state, is expressed through
the light-front variables as follows:
$$
s=\left(\sum_i k_i\right)^2 = \sum_i
\frac{ R_{i{\perp}}^2+m_i^2}{x_i},
$$
where ${\bf R}_{i{\perp}}$ and $x_i$ are
constructed according to
Eq.~(\ref{lf}). They satisfy the conservation laws similar to~Eq.~(\ref{Rx}).

\section{The integral terms in the equations~(\ref{hH})}\label{app1}
The numerators and
denominators of the kernels in the integrals in Eqs.~(\ref{hH}) are linear
functions of the scalar products
${\bf R}_{\perp} \cd {\bf R'}_{\perp}= R_{\perp}
R'_{\perp}\cos\phi'$, where ${\bf R'}_{\perp}$ is the
integration variable. We can therefore analytically integrate
over $d\phi'$, using the formulas
\begin{eqnarray*}
J_0&=&\int_0^{2\pi}\frac{d\phi'}{2\pi D(A+B\cos\phi')}=
\frac{\mbox{sign}(A)}{D\sqrt{A^2-B^2}},
\\
J_1&=&\int_0^{2\pi}\frac{\cos\phi' d\phi'}{2\pi D(A+B\cos\phi')}
=\frac{1}{DB}-\frac{A}{B}J_0.
\end{eqnarray*}
One should substitute here
\begin{eqnarray*}
A&=&{R'}^2_{\perp}(1-x)x +x'[x(x+x')m^2+R^2_{\perp}(1-x')]
\\
&-&(x+x'-1)(x'\mu^2_j+x\mu^2_{j'}),
\\
B&=&2R'_{\perp}R_{\perp}x'x,
\\
D&=&-8\pi^2 (1-x').
\end{eqnarray*}
With
$$
\eta_1=m^2x'^2+(1-x')\mu^2_{j'}+{ R'}^2_{\perp},
$$
the integral terms obtain the form
\begin{eqnarray}
i_{n}^j& =&\int_0^{\infty}  R'_{\perp} d
R'_{\perp}\int_0^{1-x} dx'\sum_{i,j'=0}^1(-1)^{j'}
\label{in}\\
&\times&
\left(c_{ni}\tilde{h}_i^{j'}
+C_{ni}\tilde{H}_i^{j'}\right),
\nonumber\\
I_{n}^j& =&\int_0^{\infty} R'_{\perp} d R'_{\perp}
\int_0^{1-x} dx'\sum_{i,j'=0}^1(-1)^{j'}
\label{In}\\
&\times&
\left(c'_{ni}\tilde{h}_i^{j'}
+C'_{ni}\tilde{H}_i^{j'}\right),\nonumber
\end{eqnarray}
for $n=0,1$. 
These integrals converge due to the PV
regularization (the sum over $j'$). The sixteen coefficients
$c$, $C$, $c'$, and $C'$ depend on $j'$. They are given below.
\begin{eqnarray*}
c_{00}&=&\frac{R'_{\perp}}{R_{\perp}\eta_1}
\Bigl\{R'_{\perp}R_{\perp}xx'J_0
\\
&+&J_1\Bigl[-{R'}^2_{\perp}(x-1)x+x'[x(-x(x'-3)
\\
&+&3x'-4)m^2 +{ R}^2_{\perp}(1-x')]
\\
&+&(x-1)(x'-1)(x'\mu^2_j+x\mu^2_{j'})\Bigr]\Bigr\},
\\
c_{01}&=&-\frac{ R'_{\perp}}{R_{\perp}}
x(2x+x'-2)J_1\ ,
\\
C_{00}&=&\frac{m^2} {R_{\perp}\eta_1}x x'[J_0
R_{\perp}(3x'-2)
\\
&+&R'_{\perp}(2-3x)J_1],
\\
C_{01}&=&\frac{x}{R_{\perp}}
 [R'_{\perp}(x-1)J_1-R_{\perp}x'J_0],
\end{eqnarray*}
\begin{eqnarray*}
 c_{10}&=&\frac{R'_{\perp}m^2}{R_{\perp}\eta_1}
x x' (x+2x'-2)J_1,
\\
c_{11}&=&-\frac{R'_{\perp}}{R_{\perp}} x
(x+x'-1)J_1\ ,
\\
C_{10}&=&-\frac{m^2}{R_{\perp}\eta_1} x x'
[R_{\perp}(1-x')J_0+ R'_{\perp}x J_1],
\\
C_{11}&=&0\ ,
\end{eqnarray*}
\begin{eqnarray*}
c'_{00}&=&\frac{R'_{\perp}}{\eta_1}  x x'
 [R'_{\perp}(3x-2)J_0+R_{\perp}(2-3x')J_1],
\\
c'_{01}&=&-\frac{R'_{\perp}}{m^2}  x
 [R'_{\perp}(x-1)J_0-R_{\perp}x'J_1],
\\
C'_{00}&=&-\frac{1}{\eta_1} \Bigl\{
{R'}^2_{\perp}(x-1)xJ_0 -R'_{\perp}R_{\perp}xx'J_1,
\\
&+&\Bigl[x\Bigl(x(x'-3)-3x'+4\Bigr)m^2
+{R}^2_{\perp}(x'-1)\Bigr]x'J_0
\\
&-&(x-1)(x'-1)(x'\mu^2_j+x\mu^2_{j'})J_0\Bigr\},
\\
C'_{01}&=&- x(2x+x'-2)J_0,
\end{eqnarray*}
\begin{eqnarray*}
c'_{10}&=&\frac{R'_{\perp}}{\eta_1}  x x'
[R'_{\perp}xJ_0 -R_{\perp}(x'-1) J_1],
\\
c'_{11}&=&0\ ,
\\
C'_{10}&=&\frac{m^2}{\eta_1}  x x' (x+2x'-2)J_0,
\\
C'_{11}&=&- x (x+x'-1)J_0.
 \end{eqnarray*}
%

\section{Coefficients in the equations~(\ref{gn2})}\label{app4}
The three-body Fock component (one fermion +
two bosons) is decomposed in four spin structures by~Eq.~(\ref{wf2})
with the coefficients
$g_{1-4}$ being scalar functions.
These coefficients  are linear combinations
(\ref{gn2}) of the functions
$\tilde{h}$ and $\tilde{H}$
which are the solution of the equations (\ref{hH}).
The coefficients of these linear combinations are
given below.

With the notation
$$
\eta_2=m^2x_2^2+(1-x_2)\mu^2_{j_2}+{R}^2_{2\perp},
$$
we have
\begin{eqnarray*}
a_{10}&=&-\frac{m({\bf R}_{1\perp}\cd {\bf
R}_{2\perp})x_2(1+x_1-x_2)} {2{R}^2_{1\perp}x_1(1-x_2) \eta_2},
\\
a_{11}&=&\frac{({\bf R}_{1\perp}\cd {\bf
R}_{2\perp})}{2m{R}^2_{1\perp}(1-x_2)}\ ,
\\
A_{10}&=&\frac{m x_2 [{R}^2_{1\perp}(1-x_2)+ ({\bf
R}_{1\perp}\cd {\bf R}_{2\perp})x_1]} {2R^2_{1\perp}x_1(1-x_2)
\eta_2},
\end{eqnarray*}

\begin{eqnarray*}
a_{20}&=&\frac{ x_2 [(1-x_2)({\bf R}_{1\perp}\cd {\bf
R}_{2\perp})+ {R}^2_{2\perp}x_1]}{2m x_1(1-x_2) \eta_2},
\\
A_{20}&=&\frac{m x_2 (1+x_1-x_2)} {2x_1(1-x_2)
\eta_2},
\\
A_{21}&=&-\frac{1}{2m(1-x_2)}, 
\end{eqnarray*}
\begin{eqnarray*}
a_{30}&=&-\frac{m^3 x_2 (1+x_1-x_2)} {2 {R}^2_{1\perp}
x_1(1-x_2)
\eta_2},
\\
a_{31}&=&\frac{m}{2  {R}^2_{1\perp}(1-x_2)},
\\
A_{30}&=&\frac{m^3 x_2 } {2 {R}^2_{1\perp}
(1-x_2)
\eta_2},
\\
a_{40}&=&\frac{m x_2}{2 x_1
\eta_2},
\end{eqnarray*}
$$
A_{11}=a_{21}=A_{31}=a_{41}=A_{40}=A_{41}=0.
$$

\section{Coefficients entering Eq.~(\ref{FF12})}\label{app3}
The three-body contributions to the form
factors $F_1$ and $F_2$ are integrals from bi-linear combinations of
the four spin components $g_{1-4}$.
The coefficients determining the three-body contribution to the
form factor $F_1$, Eq.~(\ref{F1}), have the form
\begin{eqnarray*}
C^{(1)}_{11}&=&m^4[R_{1\perp}^2 + ({\bf R}_{1\perp}\cd{\bf
\Delta})(1-x_1)],
\\
C^{(1)}_{22}&=&m^6,
\\
C^{(1)}_{44}&=&-m^2\{({\bf R}_{1\perp}\cd {\bf R}_{2\perp})^2
\\
&+& [(1-x_1)({\bf R}_{2\perp}\cd {\bf \Delta}) - x_2({\bf
R}_{1\perp}\cd {\bf \Delta})]({\bf R}_{1\perp}\cd {\bf
R}_{2\perp})
\\
&+&({\bf R}_{1\perp}\cd {\bf \Delta})R_{2\perp}^2(x_1-1)+
R_{1\perp}^2 [({\bf R}_{2\perp}\cd {\bf
\Delta})x_2-R_{2\perp}^2]\},
\\
C^{(1)}_{33}&=&\frac{1}{m^6}C^{(1)}_{11}C^{(1)}_{44},
\\
C^{(1)}_{31}&=&m^2[({\bf R}_{1\perp}\cd {\bf \Delta})({\bf
R}_{1\perp}\cd {\bf R}_{2\perp})- R_{1\perp}^2({\bf
R}_{2\perp}\cd {\bf \Delta})](x_1-1).
\end{eqnarray*}

The coefficients determining the three-body contribution
to the form factor $F_2$, Eq.~(\ref{F2}), have the form
\begin{eqnarray*}
C^{(2)}_{12}&=&-4m^4({\bf R}_{1\perp}\cd {\bf \Delta}),
\\
C^{(2)}_{41}&=&-4m^2[({\bf R}_{1\perp}\cd {\bf \Delta})({\bf
R}_{1\perp}\cd {\bf R}_{2\perp})- R_{1\perp}^2({\bf
R}_{2\perp}\cd {\bf \Delta})],
\\
C^{(2)}_{32}&=&-C^{(2)}_{41},
\\
C^{(2)}_{34}&=&\frac{4}{m^2}({\bf R}_{1\perp}\cd {\bf
\Delta})C^{(1)}_{44}.
\end{eqnarray*}
We remind that $Q^2={ \bf \Delta}^2$.\\


\end{document}